\documentclass[reprint,aps,prc,floatfix,nofootinbib,superscriptaddress]{revtex4-2} 

\pdfoutput=1
\usepackage[labelfont={small},subrefformat=parens,caption=false]{subfig}
\usepackage{amssymb}
\usepackage[utf8]{inputenc}
\usepackage[T1]{fontenc}
\usepackage{natbib}
\usepackage{xcolor}

\usepackage{mathtools} 
\usepackage{physics} 

\usepackage{bm} 
\usepackage{braket}
\usepackage{bbding} 

\usepackage{multirow}
\usepackage{tabularx}
\usepackage{colortbl}
\usepackage{booktabs} 

\usepackage{graphicx} 

\usepackage{float}
\usepackage{placeins}

\usepackage[pdfpagelabels]{hyperref} 
\usepackage[nameinlink]{cleveref}

\DeclareMathAlphabet{\curly}{OMS}{cmsy}{m}{n} 

\let\originalleft\left
\let\originalright\right
\renewcommand{\left}{\mathopen{}\mathclose\bgroup\originalleft}
\renewcommand{\right}{\aftergroup\egroup\originalright}

\newcommand{\rcite}[1]{Ref.~\cite{#1}}
\newcommand{\rscite}[1]{Refs.~\cite{#1}}
\newcommand{\CiteWithErratum}[2]{\cite{#1,*[][{ [erratum].}]#2}}

\newcommand{\unit}[1]{\ \text{#1}}

\newcommand{\ii}{i}
\newcommand{\ee}{e}

\DeclareMathOperator{\artanh}{artanh}

\DeclareMathOperator{\bigO}{\mathcal{O}}
\newcommand{\eref}[1]{=}

\DeclareMathOperator{\TrOneTwo}{Tr^{\sigma\tau}_{12}}

\DeclareMathOperator{\PureTr}{Tr^{\sigma\tau}}
\DeclareMathOperator{\Trs}{Tr^\sigma}

\newcommand{\Ps}{P^\sigma_{12}} 
\newcommand{\Pt}{P^\tau_{12}} 
\newcommand{\PNN}{P^{\sigma\tau}_{12}}

\newcommand{\vtau}{\bm{\tau}}
\newcommand{\vsigma}{\bm{\sigma}}
\renewcommand{\vnabla}{\bm{\nabla}} 

\renewcommand{\vr}{\mathbf{r}}
\newcommand{\vR}{\mathbf{R}}

\newcommand{\vx}{\mathbf{x}}

\newcommand{\vv}{\mathbf{v}}

\newcommand{\vj}{\mathbf{j}}
\newcommand{\vs}{\mathbf{s}}
\newcommand{\vrNN}{\mathbf{r}} 
\newcommand{\rNN}{r} 

\newcommand{\vRNN}{\mathbf{R}}

\newcommand{\kF}{k_\mathrm{F}}
\newcommand{\kCB}{k_\mathrm{FC}} 
\newcommand{\kFC}{\kCB}

\newcommand{\obdm}{\bm{\rho}}

\newcommand{\rhon}{\rho_\mathrm{n}}
\newcommand{\rhop}{\rho_\mathrm{p}}

\newcommand{\nmax}{{n_\mathrm{max}}}

\newcommand{\elm}[2]{$^{#2}$#1}

\graphicspath{ {./figures/} }

\crefname{table}{Table}{Tables}
\crefname{figure}{Fig.}{Figs.}
\Crefname{figure}{Figure}{Figures}
\crefname{equation}{Eq.}{Eqs.}
\Crefname{equation}{Equation}{Equations}
\crefname{section}{Sec.}{Secs.}
\Crefname{section}{Section}{Sections}

\begin{document}

\title{Comparing different density-matrix expansions for long-range pion exchange}

\author{L.~Zurek}
\email{lzurek@theorie.ikp.physik.tu-darmstadt.de}
\affiliation{Technische Universit\"at Darmstadt, Department of Physics, 64289 Darmstadt, Germany}
\affiliation{ExtreMe Matter Institute EMMI, GSI Helmholtzzentrum f\"ur Schwerionenforschung GmbH, 64291 Darmstadt, Germany}

\author{E.~A.~\surname{Coello~Pérez}}
\email{coelloperez1@llnl.gov}
\affiliation{Lawrence Livermore National Laboratory, Livermore, CA 94550, USA}
\affiliation{Technische Universit\"at Darmstadt, Department of Physics, 64289 Darmstadt, Germany}
\affiliation{ExtreMe Matter Institute EMMI, GSI Helmholtzzentrum f\"ur Schwerionenforschung GmbH, 64291 Darmstadt, Germany}

\author{S.~K.~Bogner}
\email{bogner@frib.msu.edu}
\affiliation{Facility for Rare Isotope Beams and Department of Physics and Astronomy,  \\
\mbox{Michigan State University, East Lansing, MI 48824, USA}
}

\author{R.~J.~Furnstahl}
\email{furnstahl.1@osu.edu}
\affiliation{Department of Physics, The Ohio State University, Columbus, OH 43210, USA}

\author{A.~Schwenk}
\email{schwenk@physik.tu-darmstadt.de}
\affiliation{Technische Universit\"at Darmstadt, Department of Physics, 64289 Darmstadt, Germany}
\affiliation{ExtreMe Matter Institute EMMI, GSI Helmholtzzentrum f\"ur Schwerionenforschung GmbH, 64291 Darmstadt, Germany}
\affiliation{Max-Planck-Institut für Kernphysik, Saupfercheckweg 1, 69117 Heidelberg, Germany}

\begin{abstract}
   Empirical energy density functionals (EDFs) are generally  successful in describing nuclear properties across the table of nuclides.
   But their limitations motivate using the density-matrix expansion (DME) to embed long-range pion interactions into a Skyrme functional. Recent results on the impact of the pion were both encouraging and puzzling, necessitating a careful re-examination of the DME implementation.
   Here we take the first steps, focusing on two-body scalar terms in the DME.
   Exchange energies with long-range one-pion contributions are well approximated by all DME implementations considered, with preference for variants that do not truncate at two derivatives in every EDF term.
   The use of the DME for chiral pion contributions is therefore supported by this investigation.
   For scalar-isovector energies it is important to treat neutrons and protons separately. 
   The results are found to apply under broad conditions, although self-consistency is not yet tested.
\end{abstract}

\maketitle

\section{\label{sec:intro}Introduction}

Empirical energy density functionals (EDFs) successfully describe nuclear bulk properties and some spectroscopic features throughout the known table of nuclides, except for the lightest nuclei.
Various parametrizations are available, including Skyrme, Gogny, Fayans, and relativistic functionals~\cite{Bender:2003jk,10.1088/2053-2563/aae0ed}.
Each have of order ten parameters that are determined by fits to selected nuclei.
Despite the phenomenological successes of empirical EDFs, greater accuracy is desired, e.g., for r-process nucleosynthesis~\cite{Martin:2015xql,Sprouse:2019hal} and the description of single-particle energies~\cite{Kort08SkyrmeSP}, but analyses 
suggest that the standard EDF forms have reached an accuracy limit~\cite{Kort08SkyrmeSP, Kort14UNEDF2, McDo15UncQuaEDF}.
Recasting EDFs as effective field theories (EFTs) would offer  
guidance for more accurate functionals, as well as greater control of their uncertainties and limits (e.g., toward the driplines), and would fill a gap in
the tower of EFTs describing strong interaction phenomena~\cite{Furnstahl:2019lue}.

The effort to adapt EFT methods to EDFs is complementary to the applications of chiral EFT to an expanding range of nuclei using \textit{ab initio} many-body methods~\cite{Hebeler:2015hla,Hergert:2020bxy,Tews:2020hgp}.
Chiral EFT builds on free-space internucleon interactions, while the EDFs and their extensions efficiently embody the emergent phenomena of nuclear saturation, pairing, low-lying collective excitations, and fission in a low-resolution many-body framework~\cite{Drut:2009ce,Furnstahl:2019lue}. 
However, a key question in formalizing an EFT for EDFs is the role of the pion as a long-range degree of freedom.
The pion is missing in practice from empirical EDFs, but
is it needed to reach greater accuracy for bulk properties or describing dripline physics? 
To address this question,
a semi-phenomenological hybrid approach to include pion physics based on the density-matrix expansion (DME) has been pursued in~\rscite{Gebr11DME, Stoi10DMEEDF, Dyhd17DME, Nava18DMEEDF} (see Ref.~\cite{Furnstahl:2019lue} for an overview of other approaches). 

The DME was introduced by Negele and Vautherin in their seminal papers, \rscite{Nege72DME1, Nege75DME}, as a more sophisticated alternative for approximating one-body density matrices than the simple Slater approximation~\cite{Slat50SimplHF}. 
It allows one to approximate the nonlocal one-body density matrix (OBDM) in terms of quasilocal densities by factorizing the nonlocality into universal functions. Applying it to the expression for the exchange energy in Hartree-Fock (HF) theory facilitates its calculation and clarifies how phenomenological zero-range Skyrme interactions are connected to the underlying nuclear forces. While several other DME variants have been subsequently developed (see \cref{sec:DMEvar}), a consistent extension beyond HF in many-body perturbation theory (MBPT) is not yet available~\cite{Zhan18BHFDME}.

The components of the HF-OBDM for a single particle species (neutrons or protons; identified by $q$), $\obdm_q(\vx_1, \vx_2)$, can be written in terms of self-consistent HF orbitals as
\begin{equation}
    \obdm_q(\vx_1 \sigma_1, \vx_2 \sigma_2) = \sum_{i=1}^{A_q} \phi_{q,i}^\ast(\vx_2 \sigma_2) \phi_{q,i}(\vx_1 \sigma_1) \,,
\end{equation}
with the sum running over occupied single-particle states and $A_q$ denoting the corresponding nucleon number. 
To apply a DME, the OBDM is commonly split into Hermitian scalar and vector parts~\cite{Gebr11DME}, respectively denoted by $\rho_q$ and $\vs_q$,
\begin{equation}
\label{eq:SingleSpeciesOBDMSplit}
    \obdm_q(\vx_1, \vx_2) = \frac{1}{2} \left[\rho_q(\vx_1, \vx_2) + \vs_q(\vx_1,\vx_2) \cdot \vsigma \right]
\end{equation}
with 
\begin{align}
    \rho_q(\vx_1, \vx_2) &\equiv \Trs\left[ \obdm_q(\vx_1, \vx_2) \right] \,, \\
    \vs_q(\vx_1, \vx_2) &\equiv \Trs\left[ \obdm_q(\vx_1, \vx_2) \vsigma \right] \,,
\end{align}
where the trace is in spin space only and $\vsigma$ are the spin Pauli matrices.

In the recent hybrid approach~\cite{Gebr11DME, Stoi10DMEEDF, Dyhd17DME, Nava18DMEEDF}, a simplified phase-space-averaging DME~\cite{Gebr10DME, Gebr11DME} is used to determine an EDF based on the long-range parts of free-space nucleon-nucleon (NN) and three-nucleon (3N) interactions from chiral EFT at the HF level. 
These terms are combined with a conventional Skyrme EDF whose parameters are subsequently refitted. 
The underlying idea is that the Skyrme coupling constants capture the contributions from short-range physics and correlations from higher orders in MBPT, like contact interactions in an EFT~\cite{Stoi10DMEEDF, Dyhd17DME}. 
The direct use of long-range interactions in a low-resolution EDF is justified by the observation that renormalization-group (RG) evolution only modifies the short-distance physics~\cite{Gebr11DME}.

In practice, the direct (Hartree) terms are treated exactly while the DME is applied to the exchange (Fock) contributions~\cite{Nava18DMEEDF}. The rationale for only applying the DME to the Fock contributions is twofold. First, from a purely technical standpoint, the exact treatment of the NN Hartree contribution is relatively straightforward since the direct Coulomb contribution is already treated exactly in most EDF implementations. Second, and more importantly, the falloff of the nonlocality in the Hartree and Fock contributions behaves differently~\cite{Nege72DME1}, and can result in large errors in self-consistent calculations that treat both direct and exchange terms via the DME. For the Fock contribution, the scale of nonlocality is relatively independent of direction and is set by the local Fermi momentum. In contrast, the nonlocality in the Hartree term depends strongly on direction, which is not captured in DME treatments that use angle averaging.  
EDFs obtained in this fashion have a much more involved density dependence than conventional Skyrme EDFs.

When including terms up to next-to-next-to-leading order in the chiral expansion, the hybrid-approach EDFs constructed in \rcite{Nava18DMEEDF} show a significantly improved description of experimental binding energies compared to a conventional Skyrme EDF. While these results are encouraging, the details are puzzling: first, the inclusion of 3N forces breaks this trend, and second, at leading order, which naively might be expected to have the biggest effect, no overall improvement is found. This is particularly surprising as the leading-order NN one-pion exchange constitutes the interaction with the longest range in the chiral expansion and hence is supposedly the most difficult to be correctly described by a conventional (zero-range) Skyrme pseudopotential.

There could be several reasons for this behavior, including potentially 
suboptimal choices for the DME used to determine contributions to the functional from the exchange term.
Therefore, we begin here to compare different DMEs,
focusing in particular on the accurate reproduction of the Fock energy due to long-range pion exchange.
This paper constitutes a first step towards improving the accuracy of DMEs based on chiral interactions.

At this stage, we restrict our analysis to the scalar parts of the OBDM as they contribute the most to the HF energy. 
The vector parts as well as the more involved DME choices in the 3N sector will be considered in future work. 
We also isolate the role of the DME by performing non-self-consistent tests only, with a self-consistent implementation into a full EDF planned for a later stage.

In \cref{sec:DME}, we review the different choices to be made in formulating a DME that we explore. This includes the choice of the momentum scale in DMEs, different choices for the auxiliary functions ($\Pi$ functions), expanding in single-particle vs.~center-of-mass coordinates, as well as isoscalar/isovector vs.~neutron/proton DMEs.
\Cref{sec:results} surveys results for different implementations, using different sets of orbitals for a range of closed-shell nuclei. We compare in detail the performance of the DME for the OBDM, as well as for integrated quantities, focusing on the contributions from the long-range pion exchange. In particular, our results point to the improved performance of full-square DMEs and of using DMEs for neutron/proton contributions instead of the isoscalar/isovector formulation.
Our summary and outlook are given in \cref{sec:summary}.

\section{\label{sec:DME}Density-matrix expansions}
\subsection{\label{sec:over}Overview}

For the following discussion we switch from single-particle coordinates $\vx_1$ and $\vx_2$ to relative and center-of-mass coordinates defined by
\begin{equation}
   \vr \equiv \vx_1-\vx_2 \quad\text{and}\quad \vR \equiv \frac{1}{2}(\vx_1+\vx_2) \,,
\end{equation}
and write $\rho_q(\vR; \vr)$ as a shorthand for $\rho_q(\vR + \vr/2, \vR - \vr/2) = \rho_q(\vx_1, \vx_2)$.
A naive approximation for the scalar part of the OBDM, which factorizes its nonlocality, is given by a Taylor expansion about $\vR$ truncated at order $\nmax$,
\begin{equation}
\label{eq:SchematicScalarTaylorNoAngAv}
    \rho_q(\vR; \vr) 
    \approx \sum_{n=0}^{\nmax} \frac{1}{n!} \left( \frac{\vr}{2} \cdot \vnabla_{12} \right)^n 
   \left. \vphantom{\sum_{n=0}^{\nmax}} 
    \rho_q(\vR_1,\vR_2)\right|_{\vR_1=\vR_2=\vR}\,,
\end{equation}
where $\vnabla_{12} = (\vnabla_1-\vnabla_2)$ and $\vnabla_1$ ($\vnabla_2$) acts on $\vR_1$ ($\vR_2$).
However, this approximation performs poorly at large values of $r$, for which the OBDM is expected to vanish. The latter condition can be enforced by multiplying each term of the Taylor expansion by a function $\pi_n^\nmax (k r)$ that vanishes faster than $1/r^n$ for large $r$ (using notation similar to that in Refs.~\cite{Doba03NucTh, Doba10DMEGogny}):
\begin{align}
\label{eq:SchematicScalarLocAppNoAngAv}
    \rho_q(\vR; \vr) \approx \sum_{n=0}^\nmax & \frac{\pi_n^\nmax (k r)}{n!} \left( \frac{\vr}{2} \cdot \vnabla_{12} \right)^n 
    \nonumber \\ &\null\times
    \rho_q(\vR_1,\vR_2) \Biggr|_{\vR_1=\vR_2=\vR}\,.
\end{align}
Here we have introduced the momentum scale $k$, which determines the fall-off in the off-diagonal direction of the OBDM. 
If we further impose 
\begin{equation}
    \pi_n^\nmax (x) = 1 + \bigO\left(x^{\nmax-n+1}\right) \,,
\end{equation}
the first $\nmax$ terms of the quasilocal approximation \cref{eq:SchematicScalarLocAppNoAngAv} match the first $\nmax$ terms of the Taylor series of $\rho_q(\vR; \vr)$. Specifically, the $m$th term in the Taylor expansion of \cref{eq:SchematicScalarLocAppNoAngAv} is proportional to 
\begin{equation*}
\begin{aligned}
        &(\vr \cdot \vnabla_{12})^m \rho_q(\vR_1,\vR_2) & \text{for } m \leq \nmax\,, \\
        &(r k)^{m-n} (\vr \cdot \vnabla_{12})^{n} \rho_q(\vR_1,\vR_2) & \text{for } m > \nmax\,.
\end{aligned}
\end{equation*} 

The most well-known example of such approximations is the Slater approximation~\cite{Slat50SimplHF}, which is often used in calculations of the Coulomb exchange energy~\cite{Bender:2003jk}. 
It includes only the $n=0$ term in \cref{eq:SchematicScalarLocAppNoAngAv} and is given by
\begin{equation}
    \rho_q(\vR; \vr) \approx \frac{3 j_1(\kF r)}{\kF r} \rho_q(\vR) \,,
\end{equation}
where $j_i(x)$ is a spherical Bessel function of the first kind and the local density reads
\begin{equation}
    \rho_q(\vR) \equiv \rho_q(\vR; 0)\,.
\end{equation}
The momentum scale,
\begin{equation}
\label{eq:FermiMom}
    \kF \equiv \kF^q(\vR) \equiv \left[ 3\pi^2 \rho_q(\vR) \right]^{1/3} \,,
\end{equation}
is the local density approximation for the Fermi momentum. The Slater approximation has the special feature that it becomes exact in the limit of homogeneous infinite nuclear matter (INM).

Several other approximations to the density matrix are built around the Slater approximation by adding correction terms that vanish in INM. This can be expressed nicely by regrouping certain terms in \cref{eq:SchematicScalarLocAppNoAngAv} yielding (using notation similar to Refs.~\cite{Gebr10DME, Gebr11DME})
\begin{align}
\label{eq:SchematicScalarDMENoAngAv}
    \rho_q(\vR; \vr) \approx \sum_{n=0}^{\nmax} &\frac{\Pi_n(k r)}{n!}  r_{\alpha_1}\! \cdots r_{\alpha_n}  \mathcal{P}_n^{\alpha_1 \dots \alpha_n}(\vR) \,.
\end{align}
Here and in the following, a summation over repeated Greek indices denoting spatial components is implied. The $\Pi$ functions are normalized according to $\Pi_n(0) = 1$,
and the quasilocal density combinations $\mathcal{P}_n^{\alpha_1 \dots \alpha_n}(\vR)$ are chosen such that the Taylor expansions of the exact $\rho_q(\vR; \vr)$ and of \cref{eq:SchematicScalarDMENoAngAv} agree up to order $\nmax$ (as before).
Then all terms of \cref{eq:SchematicScalarDMENoAngAv} except for the zeroth vanish in nuclear matter if $\Pi_0(x) = {3 j_1(x)}/{x}$ and $k \to \kF$ in that limit.
We refer to approximations with these properties as density-matrix expansions (DMEs) around the INM limit. 

Different DME variants differ in their choices of momentum scales $k$ and in their $\Pi$ functions. 
As the Taylor series of \cref{eq:SchematicScalarDMENoAngAv} is supposed to match the exact $\rho_q(\vR; \vr)$ only up to order $\nmax$, the higher-order terms in the $\Pi$ functions can be chosen rather unrestrictedly.
These choices can lead to significantly varying convergence behaviors with respect to $\nmax$~\cite{Carl10DMEConv}.

In general DMEs perform well at smaller $r$ values and degrade as $r$ increases, but even then, they are superior to straightforward truncations of the derivative expansion of the density matrix, \cref{eq:SchematicScalarTaylorNoAngAv}.
Additionally, one may expect that DMEs reproduce the exact OBDM better in the interior of a typical nucleus (so for small $R$) than in the nuclear surface because there the resemblance to INM is worse and the omitted higher-order terms are more relevant.

The notation of \cref{eq:SchematicScalarDMENoAngAv} has the advantage that the $\Pi$ functions do not depend on the truncation order $\nmax$ unlike the $\pi_n^\nmax$ functions used in \cref{eq:SchematicScalarLocAppNoAngAv}. However, the notation is somewhat abstract. To make it a bit more explicit we give here as an example the general expression of a second-order DME (i.e., a DME with $\nmax=2$):
\begin{align}
\label{eq:DME2ndOrderNoAngAv}
 \rho_q(\vR; \vr) & \approx \Pi_0(k \rNN)\rho_q(\vRNN)
 + \ii \Pi_1(k \rNN) r_\alpha \, j_{q, \alpha}(\vRNN)
 \nonumber \\ 
  & \quad +\frac{\Pi_2(k \rNN)}{2} r_\alpha r_\beta
 \biggl[\frac{1}{4} \nabla_\alpha \nabla_\beta \rho_q(\vRNN) - \tau_{q, \alpha \beta}(\vRNN) \nonumber \\ 
  & \qquad \null + \frac{1}{5} \delta_{\alpha \beta} k^2 \rho_q(\vRNN)\biggr]\,,
\end{align}
where the components of the current density and kinetic density tensor are given by
\begin{align}
    j_{q, \alpha}(\vR) &\equiv \left. - \frac{\ii}{2} \nabla_{12, \alpha} \rho_q(\vR_1,\vR_2)\right|_{\vR_1=\vR_2=\vR}\,, \\
    \tau_{q, \alpha\beta}(\vR) &\equiv \left. \nabla_{1,\alpha} \nabla_{2,\beta} \rho_q(\vR_1,\vR_2)\right|_{\vR_1=\vR_2=\vR}\,.
\end{align}

Finally, we note that the scalar part of the OBDM typically only has a minor dependence on the direction of the nonlocality $\vr$~\cite{Nege72DME1,Gebr10DME}. Therefore, often DMEs are formulated using an angular average with respect to $\vr$. This leads to the simpler expression
\begin{equation}
\label{eq:SchematicScalarDMEAngAv}
    \rho_q(\vR; \vr) \approx  \sum_{n=0}^{\nmax}{\vphantom{\sum}\!}' 
    \frac{\Pi_n(k r)}{n!(n+1)} r^n  \mathcal{P}_n(\vR)\,,
\end{equation}
where the prime indicates that the sum only runs over even values of $n$ (as the angular average cancels all odd-$n$ terms). 

Continuing with our example from above we obtain for a DME of order $\nmax=2$:
\begin{align}
\label{eq:DME2ndOrderAngAv}
 \rho_q(\vR; \vr) & \approx \Pi_0(k \rNN)\rho_q(\vRNN)
 +\frac{\Pi_2(k \rNN)}{6} r^2
  \nonumber \\  & \quad \null \times
  \left[\frac{1}{4} \vnabla^2 \rho_q(\vRNN) - \tau_{q}(\vRNN) + \frac{3}{5} k^2 \rho_q(\vRNN)\right]\,,
\end{align}
with the kinetic density
\begin{equation}
    \tau_{q}(\vR) \equiv \tau_{q,\alpha \alpha}(\vRNN) \,.
\end{equation}

\subsection{\label{sec:DMEvar}Considered DME variants}

Several approximations to the OBDM have been developed in the past. In this work, we explore DMEs with $\nmax \leq 2$. 
The quasilocal densities appearing in such approximations are those known from standard second-order Skyrme EDFs. Higher-order DMEs could be useful in the context of higher-order Skyrme-like EDFs~\CiteWithErratum{Carl08EDF}{Carl08Erratum} as they have the potential to be more accurate, see \rscite{Maxi01DMEord4, Carl10DMEConv} for related studies.
The considered cases are listed with their respective references in \cref{tab:DMEOverview}. Although a couple of them do not use $\Pi_0(x) = 3j_1(x)/x$, hence not reproducing the correct INM limit, see \cref{tab:DMEOverview}, we still refer to all of them as DMEs.  

Additionally, we restrict ourselves to angular-averaged DMEs as given in \cref{eq:SchematicScalarDMEAngAv,eq:DME2ndOrderAngAv}. Hence, we only list $\Pi_0(x)$ and (where applicable) $\Pi_2(x)$ in \cref{tab:DMEOverview}. 
Lifting this restriction has the potential for better accuracy, too~\cite{Carl10DMEConv}.
The DMEs considered here all use as their momentum scales either the standard local density approximation to the Fermi momentum as defined in \cref{eq:FermiMom}\footnote{See \rcite{VanV97DMEEx} for phenomenological adjustments of this momentum scale.} or an alternative introduced in \rcite{Camp78DME} which is given by
\begin{equation}
\label{eq:CampiMom}
    \kCB \equiv \kCB^q(\vR) \equiv \left\{ \frac{5}{3\rho_q(\vR)} \left[\tau_q(\vR) - \frac{1}{4} \vnabla^2 \rho_q(\vR) \right] \right\}^{1/2} \,.
\end{equation}
With the latter choice the second-order term in \cref{eq:DME2ndOrderAngAv} vanishes identically and $\Pi_2(x)$ does not need to be specified. Thus, using this momentum scale can be viewed as incorporating the second-order contribution into the zeroth-order term.
Additionally, it coincides with the regular Fermi momentum in nuclear matter, hence not changing the corresponding limit.
However, in principle, the term enclosed in square brackets in \cref{eq:CampiMom} can become negative and thus $\kFC$ imaginary. This is clearly unphysical and can lead to diverging exchange energies. In practice we find that $\kFC$ is almost always real, which has been also found in molecular systems~\cite{Lami95GauBesEx,Koeh96MolecDME}. None of the systems considered in this work produces imaginary values for $\kFC$, but for future applications one should be aware of the possibility.

\newcommand{\ts}{\textstyle}
\newcommand\myfrac[2]{\dfrac{#1}{#2\mathstrut}}
\begin{table}[t]
 \caption{
 \label{tab:DMEOverview}DME variants investigated in this work. For each DME the order $\nmax$, the expansion momentum scale $k$, and the $\Pi$ functions for the scalar parts of the OBDM are given. For the definitions of $\kF$, $\kCB$, and $G(x,Y)$ see \cref{eq:FermiMom,eq:CampiMom,eq:GaussianG}, respectively. $J_4(x)$ is the fourth (cylindrical) Bessel function of the first kind. An asterisk ($\ast$) indicates that the marked $\Pi$ function does not need to be specified as the corresponding term vanishes. The sixth column (INM) shows whether the specified DME reproduces the exact OBDM for nuclear matter. The last column (II) indicates if the DME-approximation to the OBDM obeys integrated idempotency. See text for details.}
\begin{ruledtabular}
\renewcommand\arraystretch{1.8}
\begin{tabular}{lcccccc}
DME & $\nmax$ & $k$ & $\Pi_0(x)$ & $\Pi_2(x)$ & INM & II \\ \midrule
Slater~\cite{Slat50SimplHF} & $0$ & $\kF$ & $ \myfrac{\ts 3j_1(x)}{\ts x}$ & - & \Checkmark & \Checkmark \\ 
PSA~\cite{Gebr10DME, Gebr11DME}\rule{0pt}{15pt} & $2$ & $\kF$ & $ \myfrac{\ts 3j_1(x)}{\ts x}$ & $\myfrac{\ts 3j_1(x)}{\ts x}$ & \Checkmark &  \\  
NV~\cite{Nege72DME1, Nege75DME}\rule{0pt}{15pt} & $2$ & $\kF$ & $ \myfrac{\ts 3j_1(x)}{\ts x}$ & $ \myfrac{\ts 105j_3(x)}{\ts x^3}$ & \Checkmark & \Checkmark \\ 
SVCK~\cite{Spru75DME}\rule{0pt}{15pt} & $2$ & $\kF$ & $ \myfrac{\ts 3j_1(x)}{\ts x}$ & $ \myfrac{\ts 945 j_4(x)}{\ts x^4}$ & \Checkmark & \Checkmark \\  
DT~\cite{Carl10DMEConv}\rule{0pt}{15pt} & $2$ & $\kF$ & $\myfrac{\ts 3j_1(x)}{\ts x}$ & $\exp\Bigl(-\myfrac{\ts x^2}{\ts 16}\Bigr)$ & \Checkmark & \\
CB~\cite{Camp78DME, Bhad79HODME}\rule{0pt}{15pt} & $2$ & $\kCB$ & $ \myfrac{\ts 3j_1(x)}{\ts x}$ & $\ast$ & \Checkmark & \\  
BZ~\cite{Bhad79HODME}\rule{0pt}{17pt} & $2$ & $\kCB$ & $\myfrac{\ts 96 J_4 \left(\sqrt{2}x \right)}{\ts x^4}$ & $\ast$ & & \\  
Gaussian \rule{0pt}{15pt} & $2$ & $\kCB$ & $\exp\Bigl( -\myfrac{\ts x^2}{\ts 10}\Bigr)$ & $\ast$ & & \\[-10pt] 
 \ \ \cite{Bhad79HODME, Ghos86GausAppr}\\
MG~\cite{Meye86GausOBDM}\rule{0pt}{15pt} & $2$ & $\kCB$ & $G(x, 21.5)$ & $\ast$ & & \\  
\end{tabular}
\end{ruledtabular}
\end{table}

We now proceed to give a few remarks regarding some of the considered DME variants, for more details on the variants we refer the reader to the references listed in \cref{tab:DMEOverview}:
\begin{enumerate}
\item We employ the PSA-DME in the simplified version described in \rcite{Gebr11DME} (also called INM-DME~\cite{Gebr10DME}). This is the variant that has been used in \rcite{Nava18DMEEDF} to enrich a Skyrme-like EDF with density-dependent coupling functions originating from long-range parts of chiral NN+3N interactions.

The full PSA-DME takes the anisotropy of the local momentum distribution into account, leading to a more complicated expansion momentum scale. The authors of \rscite{Gebr10DME, Gebr11DME} note that the anisotropy is especially pronounced in the surface of the nucleus and hence consider it only for the vector part of the OBDM which sharply peaks there.

The envelope of the PSA-DME $\Pi_2$ function falls off like $1/r^2$ for large r, meaning that it falls off just too slow to yield a density matrix that vanishes in the large-$r$ limit. As we will see later, this is not an issue for approximating exchange energies from finite-range forces, but it can be one in other situations.

\item The NV-DME is the ``original'' DME as formulated by Negele and Vautherin~\cite{Nege72DME1, Nege75DME}, on which 
subsequent DME developments build. For NV-DME, the authors of \rcite{Doba10DMEGogny} showed that replacing the $\Pi$ functions by exponentials having the same low-order dependence on the argument leads to almost indistinguishable results when applied to the exchange energy arising from the Gogny D1S interaction~\cite{Berg91GognyD1S}.

\item In the DT approach we use the INM limit for the model density ($\bar{\rho}^t_v$ in \rcite{Carl10DMEConv}) and set the parameter $a$ to the same value as in \rcite{Carl10DMEConv}, $a=4/\kF$. 
Note that the DT-DME, unlike the other variants, has originally been formulated without the angular averaging we use here.

\item A whole class of DMEs based on the momentum scale $\kFC$ was developed by Bhaduri and Zaifman in \rcite{Bhad79HODME} (recovering also the CB- and Gaussian DMEs). Here, we refer to the particular version rated best by them as BZ-DME.  

\item It has been argued that the Gaussian approximation is favored by information theory as it is based on the least biased phase-space distribution function subject to yielding the correct density and kinetic density distributions~\cite{Parr86GausAppI}.

In addition to the version used here, the Gaussian approximation has been developed in a form that uses the kinetic density tensor and the density's Hessian matrix instead of their scalar counterparts in \cref{eq:CampiMom}~\cite{Berk86GausAppr}, effectively amounting to using a momentum scale tensor $k_{\mathrm{FC}, \alpha\beta} (\vR)$. 

\item In the original construction~\cite{Meye86GausOBDM}, the modified Gaussian (MG) approximation uses
\begin{equation}
\label{eq:GaussianG}
   \qquad \Pi_0(x) = G(x,Y) \equiv \left( 1-\frac{x^2}{Y}\right) \ee^{-\left(\frac{1}{10}-\frac{1}{Y}\right) x^2} \,.
\end{equation}
The value of the parameter $Y>10$ then depends on the considered system and is obtained by enforcing that the approximated density matrix fulfills the integrated-idempotency constraint (as described below). This leads to the equation
\begin{multline}
\label{eq:GaussianYCond}
  \qquad  \frac{1}{A_q} \int \! \dd\vR \,  \frac{\rho_q(\vR)^2}{{\kCB^q(\vR)}^3} = 
    \frac{2}{\pi^{3/2}} \left(\frac{1}{5}-\frac{2}{Y}\right)^{3/2} 
    \\
    \null\times \left[ 1-\frac{3}{\frac{Y}{5}-2}+\frac{15}{4\left(\frac{Y}{5}-2\right)^2} \right]^{-1},
\end{multline}
which gets solved numerically for $Y$. 

We observe in our calculations that the resulting values of $Y$ do not vary much throughout the whole mass range of nuclei, thus we do not employ a specific value of $Y$ for each nucleus. Instead, we always consider a value of $Y=21.5$, which we obtained as an average over neutrons and protons in the nuclei considered in \cref{sec:results}. The resulting energies are almost indistinguishable. 

\item A modification of the Gaussian approximation similar in spirit to the MG approach has been proposed in \rcite{Lee87OBDMAppr}. 
This approximation uses $\kCB$ and 
\begin{equation}
    \Pi_0(x) = \sqrt{1 + a x^4} \exp(-x^2/10) \,,
\end{equation}
with $a$ getting determined via the integrated-idempotency constraint.
In our calculations the impact of this modification was minor, improving the results in the isovector sector but worsening them in the isoscalar case. 
For this reason we do not consider this approximation here.
\end{enumerate}

The OBDM is idempotent~\cite{Dira30Exchange,Kuma64NuclOBDM}, i.e., $\obdm_q=\obdm_q^2$, which in coordinate-space representation explicitly reads
\begin{multline}
\label{eq:Idempotency}
    \obdm_q(\vx_1\sigma_1, \vx_2\sigma_2) \\
    = \sum_{\sigma_3}\int \! \dd\vx_3 \, \obdm_q(\vx_1\sigma_1, \vx_3\sigma_3) \obdm_q(\vx_3\sigma_3, \vx_2\sigma_2) \,.
\end{multline}
In the following, we will use the matrix notation for the OBDM for brevity.
By setting $\vx_1 = \vx_2$ and noting that 
the OBDM is Hermitian,
\begin{equation}
\label{eq:OBDMdagger}
    \obdm_q(\vx_1, \vx_3)^\dagger = \obdm_q(\vx_3, \vx_1) \,,
\end{equation} 
\cref{eq:Idempotency} implies that the integrated, spin-summed version is normalized to the nucleon number
\begin{equation}
\label{eq:IntIdem}
    A_q = \frac{1}{2} \int \! \dd\vx_1 \dd\vx_3 \,  \left[ \left|\rho_q(\vx_1,\vx_3)\right|^2 + \left|\vs_q(\vx_1,\vx_3)\right|^2
    \right]
    \,.
\end{equation}

As this paper only deals with the scalar part of the OBDM we check this constraint for the considered DMEs for the case of spin-saturated nuclei where $\vs_q(\vR; \vr) \equiv 0$ in the approximation that the single-particle wave functions of spin-orbit partners are identical~\cite{Gebr10DME}. 
To this end the square of the absolute value of the scalar part of the OBDM (hereafter referred to as density-matrix square) is calculated according to the usual prescription~\cite{Koeh96MolecDME,Carl10DMEConv} that is neglecting terms of higher-than-second order [in agreement with the truncation order of \cref{eq:DME2ndOrderAngAv}]:
\begin{align}
\label{eq:Square2ndOrderAngAv}
 \left| \rho_q(\vR; \vr) \right|^2 & \approx \Pi_0(k \rNN)^2 \rho_q(\vR)^2
 +\frac{\Pi_0(k \rNN) \Pi_2(k \rNN)}{3} r^2
  \rho_q(\vRNN) \nonumber \\  & \quad \null \times
   \left[\frac{1}{4} \vnabla^2 \rho_q(\vRNN) - \tau_{q}(\vRNN) + \frac{3}{5} k^2 \rho_q(\vRNN)\right]\,.
\end{align}
We call this the truncated square of the density matrix. When calculating the square in this way only the Slater approximation as well as the NV- and SVCK-DMEs fulfill the integrated-idempotency constraint, \cref{eq:IntIdem}, exactly. In contrast, the PSA-DME violates this constraint maximally: in this case the right-hand side of \cref{eq:IntIdem} is infinite. We should also point out that the original version of the MG approximation obeys \cref{eq:IntIdem} by construction and our modification only leads to a minor deviation. 

In \cref{tab:DMEOverview} we summarize the integrated idempotency results, which for some of the considered DME variants were already given in \rscite{Bhad78DMEsolv, Koeh96MolecDME, Gebr10DME},  and also list which DMEs yield the correct INM limit.

We end this section by noting that all of the considered DMEs can be re-expressed in an orbital-free form by assuming some relation $\tau(\rho, \vnabla\rho, \dots)$ and in a completely local form by assuming $[\tau -\frac{1}{4} \vnabla^2\rho] (\rho)$, e.g., see \rcite{Lee87OBDMAppr}. This could be useful for applications to other types of EDFs than Skyrme EDFs but requires further study.

\subsection{\label{sec:square}Square of the density matrix}

In time-reversal-invariant systems the scalar part of the OBDM is real so that its Hermiticity boils down to
\begin{equation}
    \rho_q(\vR; \vr) = \rho_q(\vR; -\vr)\,,
\end{equation}
hence the current density $j_{q, \alpha}(\vR)$ vanishes~\cite{Enge75HFSkyrme}. Thus, in these cases the conventional, truncated way of calculating the density-matrix square, \cref{eq:Square2ndOrderAngAv}, which was obtained by averaging the density matrix with respect to the orientation of $\vr$ and then squaring it, has the feature of being identical to the expression one obtains from first squaring the density matrix [as given by \cref{eq:DME2ndOrderNoAngAv}] and then performing the angular average, i.e.,
\begin{equation}
    \left\langle \rho_q(\vR; \vr) \right\rangle_{\Omega_\vr} ^2 = \left\langle \rho_q(\vR; \vr)^2 \right\rangle_{\Omega_\vr} \,.
\end{equation}
Here $\langle \dots \rangle_{\Omega_\vr}$ indicates averaging over the direction of $\vr$.
However, for certain DME variants \cref{eq:Square2ndOrderAngAv} also possesses the undesirable characteristic of yielding a negative-valued square for some values of $\vR$ and $r$. 

\begin{figure}[htb]
\includegraphics[width=1\linewidth]{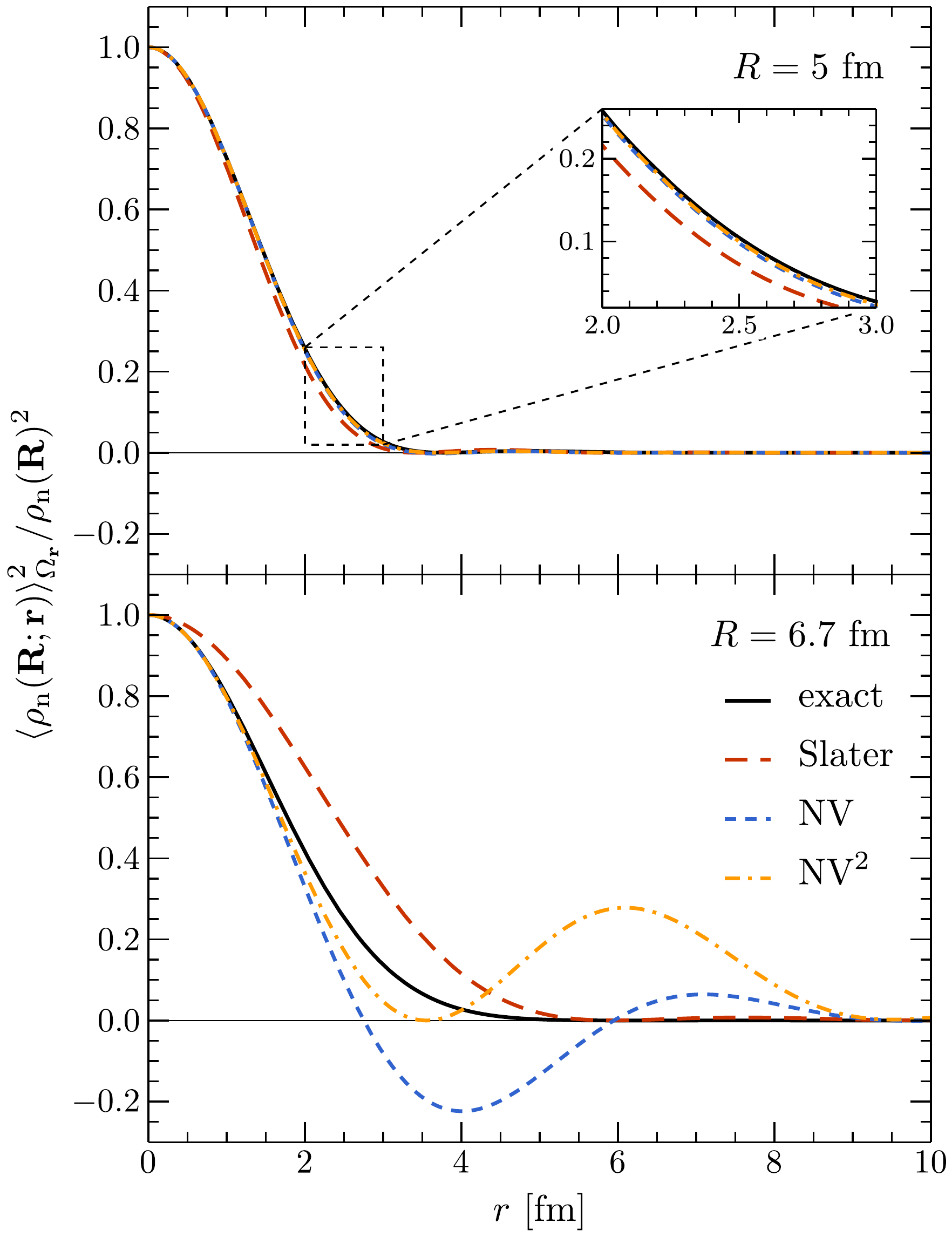}
\caption{\label{fig:rho2}Normalized density-matrix square for two values of $R$ in $^{132}$Sn for different DMEs. The underlying orbitals are obtained from a self-consistent HF calculation with the SLy4-EDF.}
\end{figure}

\Cref{fig:rho2} contains an example of such behavior: We show the density-matrix square for neutrons in $^{132}$Sn as a function of the nonlocality $r$ for two values of $R$, 5~fm (just in the surface of the nucleus, see \cref{fig:DensityDistributions}) and 6.7~fm (quite far into the surface of the nucleus). The underlying single-particle orbitals are generated from a self-consistent HF calculation using the SLy4-EDF~\CiteWithErratum{Chab98SLy}{Chab98Erratum}. 
In addition to the exact square in solid black, \cref{fig:rho2} includes the Slater approximation and the NV-DME as defined in \cref{tab:DMEOverview}. For $R=6.7\unit{fm}$, where the second-order correction is much larger (relative to the zeroth-order term), the NV-DME significantly underestimates the value of the square and becomes negative for $2.8\unit{fm} \lesssim r \lesssim 5.9\unit{fm}$.

Therefore, we additionally employ an alternative approach for squaring the density matrix. It was briefly investigated in \rcite{Camp78DME} and consists in considering the full square of the angle-averaged density matrix, \cref{eq:DME2ndOrderAngAv}, (hence the approximation cannot get negative): 
\begin{align}
\label{eq:FullSquare2ndOrderAngAv}
 \rho_q(\vR; \vr) ^2 & \approx \biggl\{ \Pi_0(k \rNN)\rho_q(\vRNN)
 +\frac{\Pi_2(k \rNN)}{6} r^2
 \nonumber \\  & \quad \null \times \left[\frac{1}{4} \vnabla^2 \rho_q(\vRNN) - \tau_{q}(\vRNN) + \frac{3}{5} k^2 \rho_q(\vRNN)\right] \biggr\}^2 \,.
\end{align}
Whereas this equation is not in agreement with the truncation order of \cref{eq:DME2ndOrderAngAv} and thus contains some but not all of the fourth-order terms, it effectively shifts the region where the DME approximation is poor to larger $r$ values compared to the previously applied truncated squaring prescription.
This can be seen in the lower panel of \cref{fig:rho2}, where this approach is labeled as NV$^2$. Such behavior can turn out useful when approximating expressions where the large-$r$ behavior is damped, e.g., exchange energies from finite-range forces as considered in \cref{sec:results}. 
In the following we refer to the same treatment of the square for \mbox{PSA-,} \mbox{SVCK-,} and DT-DMEs as \mbox{PSA$^2$-,} \mbox{SVCK$^2$-,} and DT$^2$-DMEs, respectively. The other investigated DME variants have no contribution from $\Pi_2$, hence \cref{eq:Square2ndOrderAngAv} and \cref{eq:FullSquare2ndOrderAngAv} yield identical results in those cases.

We note that
this treatment of the square makes the DMEs no longer fulfill the integrated idempotency.
In addition, the statement that squaring and angular averaging commute is no longer true:
\begin{equation}
    \left\langle \rho_q(\vR; \vr) \right\rangle_{\Omega_\vr}^2 \neq \left\langle \rho_q(\vR; \vr)^2 \right\rangle_{\Omega_\vr} \,.
\end{equation}
In the particular case of the PSA-DME the truncated squaring approach yields a density-matrix square that vanishes for large $r$, but the PSA$^2$-DME does not (see also the related remark in \cref{sec:DMEvar}). 
We also note that while still being constructed from the standard quasilocal Skyrme densities, the full-square DME variants lead to EDFs with more than two derivatives in some terms (as do all DMEs with the $\kFC$ momentum scale).

\subsection{\label{sec:coords}Expansion coordinates}

Up until now we have expanded the density matrices in the relative coordinate $\vr$ around the center of mass $\vR$. However, other choices are possible. It is useful to choose the expansion point to be located on the line connecting the two positions of interest, $\vx_1$ and $\vx_2$. Then, the nonlocality can be fully expressed in terms of only one coordinate, $\vr$. 
This is generally not possible when dealing with 3N forces, e.g., see \rcite{Dyhd17DME}, and constitutes one of the reasons why applying a DME for such interactions is much more involved.
Here we deal only with NN forces and are thus able to express the nonlocality only in terms of $\vr$. We refer to the general expansion point in between $\vx_1$ and $\vx_2$ as 
\begin{equation}
    \vv_a \equiv a \vx_1 + (1-a) \vx_2
\end{equation}
where $a \in [0,1]$ determines the exact expansion point. For $a=1/2$ one recovers the center of mass $\vR$ used in the previous sections and for $a=0$ the expansion is about the position of the second particle.
For this expansion point the analogous expression to \cref{eq:DME2ndOrderNoAngAv} for a DME of order $\nmax=2$ reads 
\begin{align}
\label{eq:DME2ndOrderGeneralNoAngAv}
 \rho_q(\vv_a; \vrNN) &\equiv \rho_q(\vv_a +(1-a) \vrNN, \vv_a -a \vrNN) \nonumber \\
 & \approx \Pi_0(k \rNN)\rho_q(\vv_a)
 \nonumber \\ 
  & \quad 
  + \Pi_1(k \rNN) r_\alpha \left[ \left( \frac{1}{2}-a \right) \nabla_\alpha \rho(\vv_a) + \ii j_{q, \alpha}(\vv_a) \right]
 \nonumber \\ 
  & \quad +\frac{\Pi_2(k \rNN)}{2} r_\alpha r_\beta
 \biggl[ \left( \frac{1}{2}-a+a^2 \right) \nabla_\alpha \nabla_\beta \rho_q(\vv_a)
 \nonumber \\ 
  & \qquad + (1-2a) \ii \nabla_\alpha j_{q,\beta}(\vv_a) 
  - \tau_{q, \alpha \beta}(\vv_a) \nonumber \\ 
  & \qquad \null + \frac{1}{5} \delta_{\alpha \beta} k^2 \rho_q(\vv_a)\biggr]\,.
\end{align}
Averaging over the direction of $\vr$ yields
\begin{align}
\label{eq:DME2ndOrderGeneralAngAv}
 \rho_q(\vv_a; \vr) & \approx \Pi_0(k \rNN)\rho_q(\vv_a)
 \nonumber \\  & \quad 
 +\frac{\Pi_2(k \rNN)}{6} r^2
  \biggl[ \left( \frac{1}{2}-a+a^2 \right) \vnabla^2 \rho_q(\vv_a) 
  \nonumber \\ 
  & \qquad
  + (1-2a) \ii \vnabla \cdot \vj_q(\vv_a) - \tau_{q}(\vv_a) 
  \nonumber \\ 
  & \qquad
  \null + \frac{3}{5} k^2 \rho_q(\vv_a)\biggr]\,,
\end{align}
which simplifies to \cref{eq:DME2ndOrderAngAv} when $a=1/2$.

For $a \neq 1/2$, angular averaging and squaring do not commute even for time-reversal-invariant systems. Both for the truncated and the full squaring prescription,
\begin{equation}
\label{eq:AvSqCommuteGeneral}
    \left\langle \rho_q(\vv_a; \vr)  \right\rangle_{\Omega_\vr} \left\langle \rho_q(\vv_a; \vr)^\ast \right\rangle_{\Omega_\vr} \neq \left\langle \left| \rho_q(\vv_a; \vr) \right|^2 \right\rangle_{\Omega_\vr} \,,
\end{equation}
as a term proportional to $(\vnabla \rho_q)^2$ is missing on the left-hand side of \cref{eq:AvSqCommuteGeneral}.

Nevertheless, \cref{eq:DME2ndOrderGeneralAngAv} and an accordingly adjusted momentum scale $\kF^q(\vv_a)$ were used in \rcite{Koeh96MolecDME} in time-reversal-invariant molecular systems with the NV-DME, and $a=0$ was found to lead to a much improved reproduction of the exact Coulomb exchange energies compared to the usual $a=1/2$ choice.
An optimization routine gave basically the same value ($a=0.00638$) as the best fit for their considered systems~\cite{Koeh96MolecDME}.
While we are able to reproduce a similar behavior in our test systems when using the Coulomb interaction, we do not see this improvement for one-pion exchange, see \cref{sec:DepInteraction} for details.

Moreover, it is not clear how to extend the DME variants that use $\kFC$ to $a=0$ because for $\Pi_2$ to not contribute one needs an adjusted momentum scale, 
\begin{equation}
    \tilde{k}_\mathrm{FC}^q(\vx_2) \equiv \left\{ \frac{5}{3\rho_q(\vx_2)} \left[\tau_q(\vx_2) - \frac{1}{2} \vnabla^2 \rho_q(\vx_2) \right] \right\}^{1/2} \,
\end{equation}
(note the prefactor $1/2$ instead of $1/4$ in front of $\vnabla^2 \rho_q$). This $\tilde{k}_\mathrm{FC}^q$ is often imaginary~\cite{Koeh96MolecDME}, which is unphysical and can lead to diverging exchange energies.
For these reasons we only consider DMEs about $\vR$ in the following (except for \cref{sec:DepInteraction} as noted).

\section{Results and discussion}\label{sec:results} 

We now proceed to apply the different DME variants discussed in \cref{sec:DMEvar,sec:square} to the nonlocal densities in the exchange energy arising from a local NN interaction in coordinate space. This energy is given by~\cite{Bogn09DME, Dyhd17DME}
\begin{align}
\label{eq:FockLocalNN}
    W_\text{ex} &= -\frac{1}{2} \TrOneTwo \int \! \dd\vR\, \dd\vr \, \obdm^{(1)}(\vR; -\vr) \obdm^{(2)}(\vR; \vr) 
    \nonumber \\ &\quad \null \times
    \braket{ \vr | V^{(1 \otimes 2)} | \vr } \PNN \,,
\end{align}
where the index 1 (2) denotes on which part of the two-body product space the OBDMs and the potential $V$ act, i.e., 1 (2) refers to the spin and isospin space of ``particle 1'' (``particle 2''), $\TrOneTwo$ denotes a trace over the whole product space, and 
\begin{equation}
  \PNN \equiv \Ps \Pt = \frac{1 + \vsigma_1 \cdot \vsigma_2}{2} \frac{1 + \vtau_1 \cdot \vtau_2}{2}   
\end{equation}
is the two-particle spin and isospin exchange operator, with isospin Pauli matrices $\vtau$.
The OBDMs in \cref{eq:FockLocalNN} are those of the whole system and can be split similarly to \cref{eq:SingleSpeciesOBDMSplit},
\begin{align}
\label{eq:OBDMPartitioning}
    \obdm(\vR; \vr) &= \frac{1}{4} [\rho_0(\vR; \vr) + \rho_1(\vR; \vr) \tau_z 
    \nonumber \\ & \quad \null + \vs_0(\vR; \vr) \cdot \vsigma + \vs_1(\vR; \vr) \cdot \vsigma \tau_z ] \,,
\end{align}
where we assumed that the single-particle states do not mix neutrons and protons. The scalar-isoscalar, scalar-isovector, vector-isoscalar, and vector-isovector parts are given by 
\begin{align}
    \rho_0(\vR; \vr) &\equiv \PureTr\left[ \obdm(\vR; \vr) \right] \,,\\
    \rho_1(\vR; \vr) &\equiv \PureTr\left[ \obdm(\vR; \vr) \tau_z \right] \,,\\
    \vs_0(\vR; \vr) &\equiv \PureTr\left[ \obdm(\vR; \vr) \vsigma \right] \,,\\
    \vs_1(\vR; \vr) &\equiv \PureTr\left[ \obdm(\vR; \vr) \vsigma \tau_z \right] \,.
\end{align}
Isoscalar and isovector quantities are sums and differences of the corresponding neutron and proton quantities, e.g.,
\begin{align}
\label{eq:IsoscalarSum}
    \rho_0(\vR; \vr) &= \rhon(\vR; \vr) + \rhop(\vR; \vr) \,, \\
\label{eq:IsovectorDiff}
    \rho_1(\vR; \vr) &= \rhon(\vR; \vr) - \rhop(\vR; \vr) \,,
\end{align}
which are treated separately when expanded with a DME.
After breaking up the nonlocal densities as in \cref{eq:OBDMPartitioning} the exchange energy reads
\begin{align}
\label{eq:FockLocalNNParts}
    W_\text{ex} &= -\frac{1}{32} \TrOneTwo \int \! \dd\vR\, \dd\vr \, 
    \Bigl[\rho_0(\vR; -\vr) + \rho_1(\vR; -\vr) \tau_z^{(1)} 
    \nonumber \\ & \qquad \null + \vs_0(\vR; -\vr) \cdot \vsigma^{(1)} + \vs_1(\vR; -\vr) \cdot \vsigma^{(1)} \tau_z^{(1)} \Bigr]
    \nonumber \\ & \quad \null \times \Bigl[\rho_0(\vR; \vr) + \rho_1(\vR; \vr) \tau_z^{(2)} + \vs_0(\vR; \vr) \cdot \vsigma^{(2)} 
    \nonumber \\ & \qquad \null + \vs_1(\vR; \vr) \cdot \vsigma^{(2)} \tau_z^{(2)} \Bigr]
    \braket{ \vr | V^{(1 \otimes 2)} | \vr } \PNN \,.
\end{align}
Depending on the spin and isospin structure of the interaction, different bilinears of the OBDM parts survive in \cref{eq:FockLocalNNParts} after carrying out the traces. 

To test the different DMEs we insert these approximations into \cref{eq:FockLocalNNParts} and compare the resulting energies to the exact exchange energy. Before we can do that we need to specify both the system (which enters the OBDMs) and the interaction. Let us start with discussing the latter.
As stated earlier, we restrict ourselves to NN interactions in this work because the inclusion of 3N forces involves dealing with two relative coordinates in the OBDMs (instead of one), which means that even more approximations and choices need to be considered. This study will be carried out in future work.

DMEs are naturally formulated in coordinate space. Thus, using them together with momentum-space interactions requires explicitly evaluating a Fourier transform (e.g., see \rscite{Bogn09DME,Holt11ChiEDF}) which hinders linking observations with the form of the $\Pi$ functions. For coordinate-space interactions a Fourier transform is not necessary and the analysis is more straightforward. Therefore, we consider only interactions formulated in coordinate space. 

As DMEs are less accurate for large values of the relative distance $r$, a good description of the exchange energy arising from long-range interactions is particularly challenging. The interactions used in \rcite{Nava18DMEEDF} to enrich a Skyrme EDF are determined from chiral EFT. In this scheme, the interaction with the longest range is one-pion exchange, which appears already at leading order (LO) in the chiral expansion, meaning that it should be particularly relevant according to the underlying power counting. 
Investigating one-pion exchange is also interesting because the inclusion of this term in \rcite{Nava18DMEEDF} did not improve the functional's reproduction of experimental binding energies (unlike for higher order, shorter-range terms). 

The one-pion exchange piece with the longest range is described by a central Yukawa interaction, which in coordinate space reads:  
\begin{equation}
\label{eq:WSMatrix}
    \braket{ \vr | V^{(1 \otimes 2)} | \vr } = W_S^{\mathrm{LO}}(r) \vsigma_1 \cdot \vsigma_2 \vtau_1 \cdot \vtau_2 \,,
\end{equation}
with the radial dependence 
\begin{equation}
    W_S^{\mathrm{LO}}(r) \equiv  \frac{m_\pi^3}{12\pi} \left( \frac{g_{A}}{2 F_\pi} \right)^2 \frac{\ee^{-m_\pi r}}{m_\pi r} \,,
\end{equation}
where we use $g_A = 1.29$, $F_\pi = 92.4 \unit{MeV}$, and $m_\pi = 138.03 \unit{MeV}$ for the axial-vector coupling constant, the pion decay constant, and the pion mass, respectively~\cite{Geze14long}.
To regularize the interaction it is multiplied with a local regulator function $f(r)$,
\begin{equation}
\label{eq:RegReplace}
    W_S^{\mathrm{LO}}(r) \to W_S^{\mathrm{LO}}(r) f(r) \,.
\end{equation}
While other coordinate-space regulator forms are available, e.g., see \rcite{Epel15improved}, we choose here~\cite{Geze13QMCchi, Geze14long, Dura18DoubFold}
\begin{equation}
\label{eq:Regulator}
f(r) = 1 - \exp( - \frac{r^4}{R_0^4} ) \,,
\end{equation}
where the spatial cutoff $R_0$ specifies up to which value of $r$ the short-distance part of the potential is smoothly cut off. We first consider $R_0 = 1.2\unit{fm}$.
While regulators are not needed at the HF level, they suppress large short-distance contributions~\cite{Tews16QMCPNM, Dyhd16Regs} that would otherwise have to be absorbed into the Skyrme parameters and enable us to smoothly turn on the long-distance interactions.

The tensor part of one-pion exchange has a shorter range than the central piece and its exchange energy involves only the vector part of the OBDM, so we do not consider it here. 
Applying a DME to the short-range piece of one-pion exchange (whether described by a smeared-out delta function or an actual one) works very well because of its short range. 
In a scheme where a proper delta function is used all DME variants even yield the same (exact) functional with density-independent couplings as in a Skyrme EDF.  

Inserting \cref{eq:WSMatrix} into \cref{eq:FockLocalNNParts} yields for the $W_S^{\mathrm{LO}}$ exchange energy
\begin{align}
    W_\text{ex} &= -\frac{1}{8} \int \! \dd\vR\, \dd\vr \, \bigl[ 9 |\rho_0(\vR; \vr)|^2 - 3 |\rho_1(\vR; \vr)|^2
    \nonumber \\ & \quad
    \null - 3 |\vs_0(\vR; \vr)|^2 + |\vs_1(\vR; \vr)|^2 \bigr] W_S^{\mathrm{LO}}(r) f(r) \,.
\end{align}
We consider the first two terms (which depend on the scalar parts of the OBDM) and refer to them as the
scalar-isoscalar energy $W_0$,
\begin{equation}
\label{eq:IsoscalarEnergy}
    W_0 \equiv -\frac{9}{8} \int \! \dd\vR\, \dd\vr \,  |\rho_0(\vR; \vr)|^2 
    W_S^{\mathrm{LO}}(r) f(r) \,,
\end{equation}
and scalar-isovector energy $W_1$,
\begin{equation}
\label{eq:IsovectorEnergy}
    W_1 \equiv \frac{3}{8} \int \! \dd\vR\, \dd\vr \,  |\rho_1(\vR; \vr)|^2 
    W_S^{\mathrm{LO}}(r) f(r)\,.
\end{equation}
The question we want to investigate now is how well do the different DME variants of \cref{sec:DMEvar,sec:square} approximate these energies.
 
\begin{figure}[t]
\includegraphics[width=\linewidth]{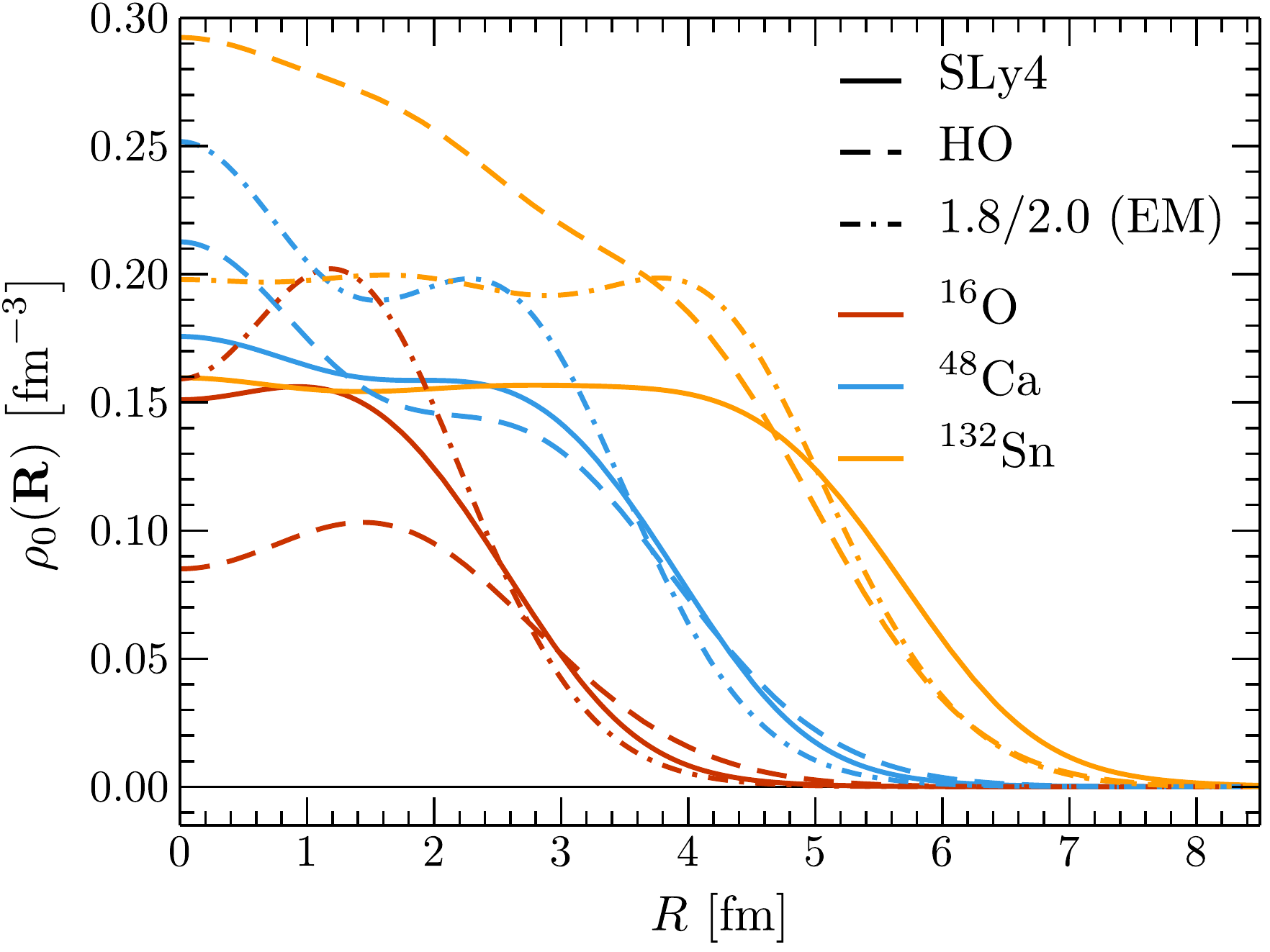}
\caption{\label{fig:DensityDistributions}Isoscalar density distributions of selected closed-shell nuclei. Solid lines correspond to orbitals from a self-consistent HF calculation with the SLy4-EDF, dashed lines correspond to orbitals from an isotropic HO with $\hbar\omega=10\unit{MeV}$, and dash-dotted lines correspond to orbitals obtained from a self-consistent HF calculation with the 1.8/2.0 (EM) interaction~\cite{Hebe11fits}.}
\end{figure}

In this work, we compare different DMEs with the same single-particle orbitals generated from a self-consistent HF calculation employing the SLy4 parametrization of the Skyrme EDF~\CiteWithErratum{Chab98SLy}{Chab98Erratum} without pairing. This enables a clean comparison, but we point out that the orbitals used are not self-consistent with the EDF and DME. The HF equations are solved using the code \textsc{hfbrad}~\cite{Benn05HFBRAD}, which works directly on a spherical coordinate-space grid. 
The step size is set to $0.1\unit{fm}$. Reducing the step size to $0.025\unit{fm}$ changes the obtained total energies of the HF calculation at most in the per-mill regime. This precision is sufficient for the present application. 
We made sure the code and the implementation of the outputted orbitals into our DME routines work as intended by comparing against results obtained with orbitals from \textsc{hosphe}~\cite{Carl10hosphe} and \textsc{hfodd}~\cite{Schu17hfodd}.
The DME implementations themselves were benchmarked against the second-order results of \rcite{Carl10DMEConv}, the LO results of \rcite{Dyhd17DME}, and the one-pion-exchange Fock expressions of \rscite{Kais03MedDME,Kais10MedDME}.
We consider in total 11 closed-shell nuclei, ranging from light to heavy and from $N=Z$ to very asymmetric: \elm{O}{16}, \elm{O}{24}, \elm{Ca}{40}, \elm{Ca}{48}, \elm{Ca}{54}, \elm{Ni}{56}, \elm{Ni}{60}, \elm{Zr}{80}, \elm{Sn}{100}, \elm{Sn}{132}, and \elm{Pb}{208}.
All of these nuclei are closed-shell, hence their ground states are treated as being time-reversal invariant and $\Pi_1$ does not contribute even without the angular-average approximation.
For three example nuclei the isoscalar density distributions are displayed as solid lines in \cref{fig:DensityDistributions}. 

In the following subsections we first consider the scalar-isoscalar energy in \cref{sec:Isosca} and then the scalar-isovector energy in \cref{sec:Isovec}, followed by an analysis regarding the use of a single species-independent momentum scale in \cref{sec:IsoscaMom}. In \cref{sec:DepOrbitals,sec:DepInteraction} we discuss the dependence of the results on the considered orbitals and 
on the employed interaction and expansion coordinates, respectively.
We finish each subsection with a summary of the main take-away points.

\subsection{\label{sec:Isosca}Scalar-isoscalar energy} 

\begin{figure*}[t]
\includegraphics[width=1\linewidth]{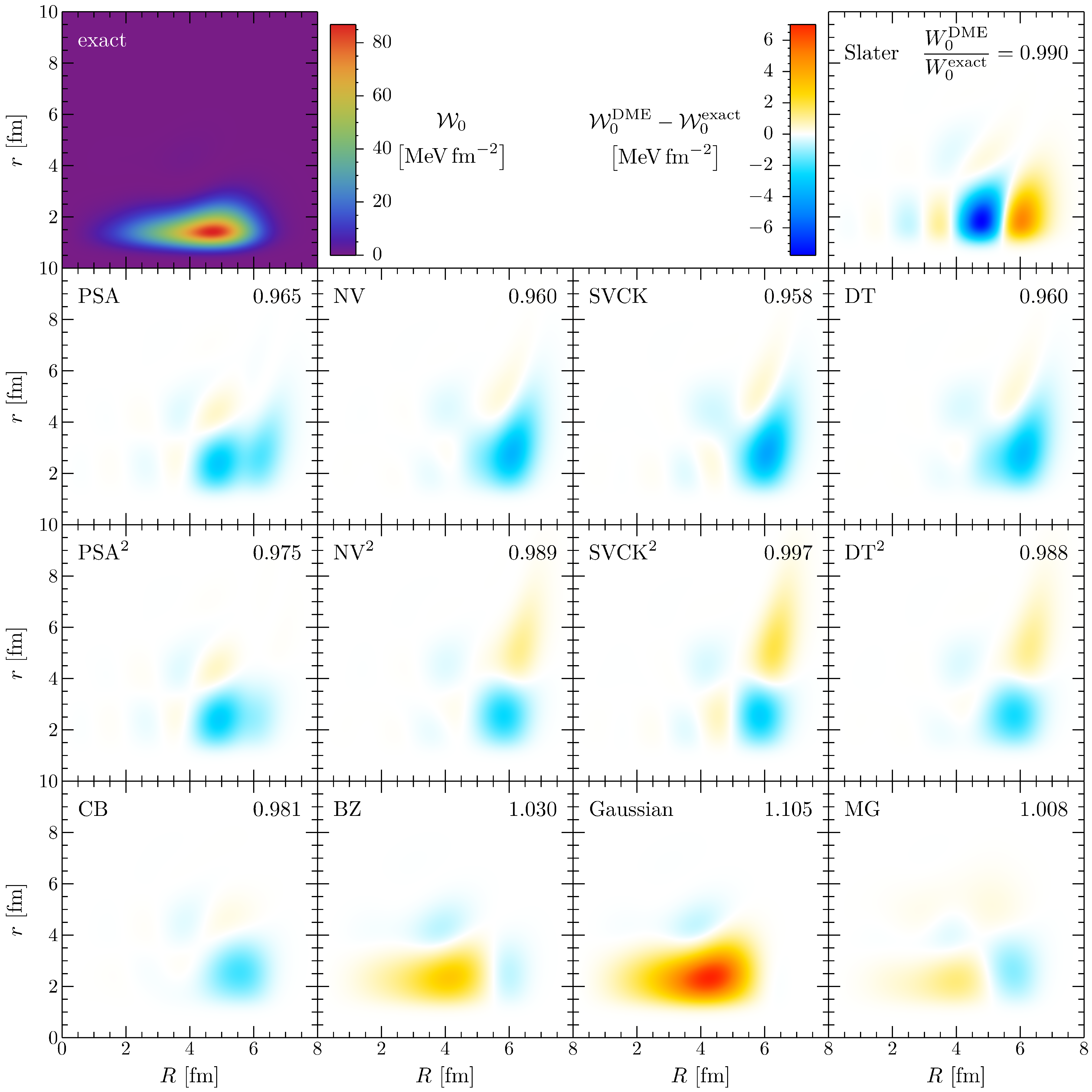}
\caption{\label{fig:0Integrand-Sn132}First panel: exact scalar-isoscalar exchange-energy integrand obtained for a regularized Yukawa interaction in $^{132}$Sn. Other panels: differences of DME approximations of this integrand and the exact integrand itself. In every difference panel the value of the ratio of the DME-approximated energy and the exact energy is shown in the top right corner. The underlying orbitals are obtained from a self-consistent HF calculation with the SLy4-EDF.}
\end{figure*}

We begin by considering the scalar-isoscalar Yukawa exchange-energy integrand $\mathcal{W}_0$, defined as
\begin{align}
\label{eq:0Integrand}
    \mathcal{W}_0 (R,r) &\equiv \frac{9}{8} \int \! \dd\Omega_{\vR} \, \dd\Omega_{\vr} \, R^2 r^2 |\rho_0(\vR; \vr)|^2
    W_S^{\mathrm{LO}}(r) f(r)\,,
\end{align}
and pick \elm{Sn}{132} as our first test case. The \elm{Sn}{132} isoscalar (matter) density distribution is shown in \cref{fig:DensityDistributions}. The exact integrand $\mathcal{W}_0$, displayed in the first panel of \cref{fig:0Integrand-Sn132}, has the largest contributions at about $r \approx 1.5\unit{fm}$ over the whole $R$ range, which mostly reflects the nature of the regularized interaction, and peaks at $R \approx 4.7\unit{fm}$, which is close to the peak of $R^2 \rho_0(\vR)^2$ at $R \approx 4.6\unit{fm}$, the expected peak position for an exactly separable OBDM. 
These features are rather well reproduced by the integrands that are obtained when replacing $|\rho_0(\vR; \vr)|^2$ in \cref{eq:0Integrand} by its different DME approximations. Therefore, we do not show the DME integrands themselves but instead their differences to the exact integrand. 
They are depicted in the other panels of \cref{fig:0Integrand-Sn132}: the zeroth-order Slater approximation in the top-right panel, DMEs using $\kF$ with the common truncated-square approach, \cref{eq:Square2ndOrderAngAv}, in the second row, and with full squares, \cref{eq:FullSquare2ndOrderAngAv}, in the third row, and DMEs with $\kFC$ in the last row. The same order and grouping is used in the other figures below.

Several trends are clearly visible from the integrand differences in \cref{fig:0Integrand-Sn132}: 
The second-order DMEs (besides the Gaussian approximation) locally reproduce the exact integrand significantly better than the zeroth-order Slater approach, highlighting the improvement due to the inclusion of higher-order terms. In particular, the region where relevant deviations first occur gets shifted from $r \approx 1\unit{fm}$ in the Slater case to $r \approx 1.5\unit{fm}$ for the other DMEs.
In all cases the largest differences arise close to or in the surface of the nucleus. This can probably be attributed to the larger relevance of missing higher-order terms compared to the situation in the interior of the nucleus. 
Comparing the second and third rows of panels in \cref{fig:0Integrand-Sn132} reveals that the additional term in the full-square DMEs is particularly relevant in the surface where it flips the sign of the differences for $r \gtrsim 4 \unit{fm}$. Interestingly, this is not always an improvement locally but the global scalar-isoscalar energy $W_0$ is always closer to the exact result for the full square than for the truncated-squares approach due to (possibly fortuitous) cancellations in the former case. We provide the ratio of the DME-approximated $W_0$ and the exact counterpart in the top-right corner of each panel.

Regarding these global energies, all considered DME variants approximate the exact values remarkably well with SVCK$^2$- and MG-DMEs performing best: both yield values that deviate less than 1\% from the exact result. 
Somewhat surprisingly, the Slater approximation follows next despite the inferior quality in local reproduction of the integrand. Again, this can be attributed to cancellations of regions of overestimation and underestimation. 
We also note that while the Gaussian approximation overestimates the integrand throughout the nuclear interior, yielding the worst energy reproduction, it provides an extremely good description of the integrand in the surface.

\begin{figure}[t]
\includegraphics[width=1\linewidth]{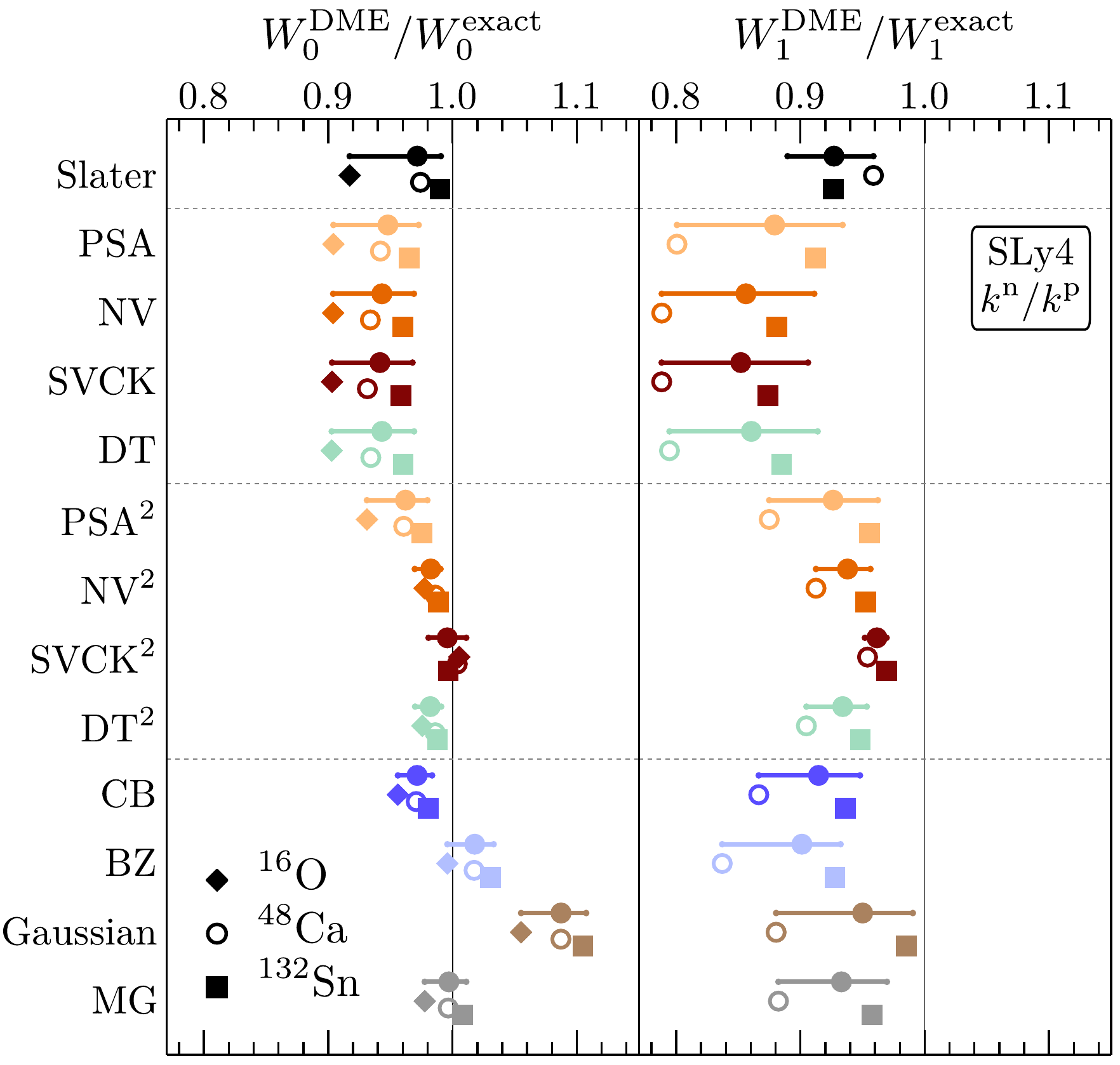}
\caption{\label{fig:Ratios-np-SLy4}Ratios of DME-approximated and exact exchange-energy contributions for a regularized Yukawa interaction. 
For every DME variant the average over a set of nuclei is shown together with a bar ranging from the smallest to the largest ratio observed. Below each bar, the results for selected nuclei are given. 
The underlying orbitals are obtained from a self-consistent HF calculation with the SLy4-EDF. 
The left panel shows the results for the scalar-isoscalar contributions, where 11 closed-shell nuclei are considered (see text); the right panel shows the results for the scalar-isovector contributions (from 6 closed-shell nuclei). 
In both panels the DMEs are used with separate momentum scales for neutrons and protons.}
\end{figure}

In other nuclei, the results are very similar. This can be seen from the left panel in \cref{fig:Ratios-np-SLy4} where we show for each DME variant the average ratio of approximated ($W_0^\text{DME}$) and exact ($W_0^\text{exact}$) scalar-isoscalar energies over the 11 test nuclei and a bar that ranges from the smallest to the largest ratio observed. Underneath each bar the values of three individual nuclei (corresponding to the density distributions shown in \cref{fig:DensityDistributions}) are highlighted, showing that smaller ratios (typically corresponding to worse energy reproductions) almost always occur for lighter nuclei.

As before, all ratios are notably close to unity and the full-square variants of the DMEs reproduce the exact energies better than the corresponding truncated-square versions. Additionally, the spread of the ratios is smaller for the full squares. 
Again, the reproduction is particularly good for SVCK$^2$- and MG-DME and on average it is worst for the Gaussian approximation.

We find almost identical, though slightly worse, results when approximating the NV-DME $\Pi$ functions with exponentials as proposed in \rcite{Doba10DMEGogny}. This approximation could be useful for implementations in numerical EDF codes.
For the MG-DME the results are almost indistinguishable when using the average value of the parameter $Y=21.5$ as done here and when using specific values for each nucleus based on the integrated-idempotency constraint as described in \cref{sec:DMEvar}.

In summary, using DMEs to approximate the scalar-isoscalar energies $W_0$ works remarkably well for the considered closed-shell systems and the investigated Yukawa interaction. The dependence on orbitals and interaction is investigated in \cref{sec:DepOrbitals,sec:DepInteraction}, respectively.
Refined improvement from few-percent accuracy for some DME variants to the 1\% level can be realized by switching to full-square DMEs or other variants, in particular to SVCK$^2$- and MG-DME.

\subsection{\label{sec:Isovec}Scalar-isovector energy} 

\begin{figure*}[t]
\includegraphics[width=1\linewidth]{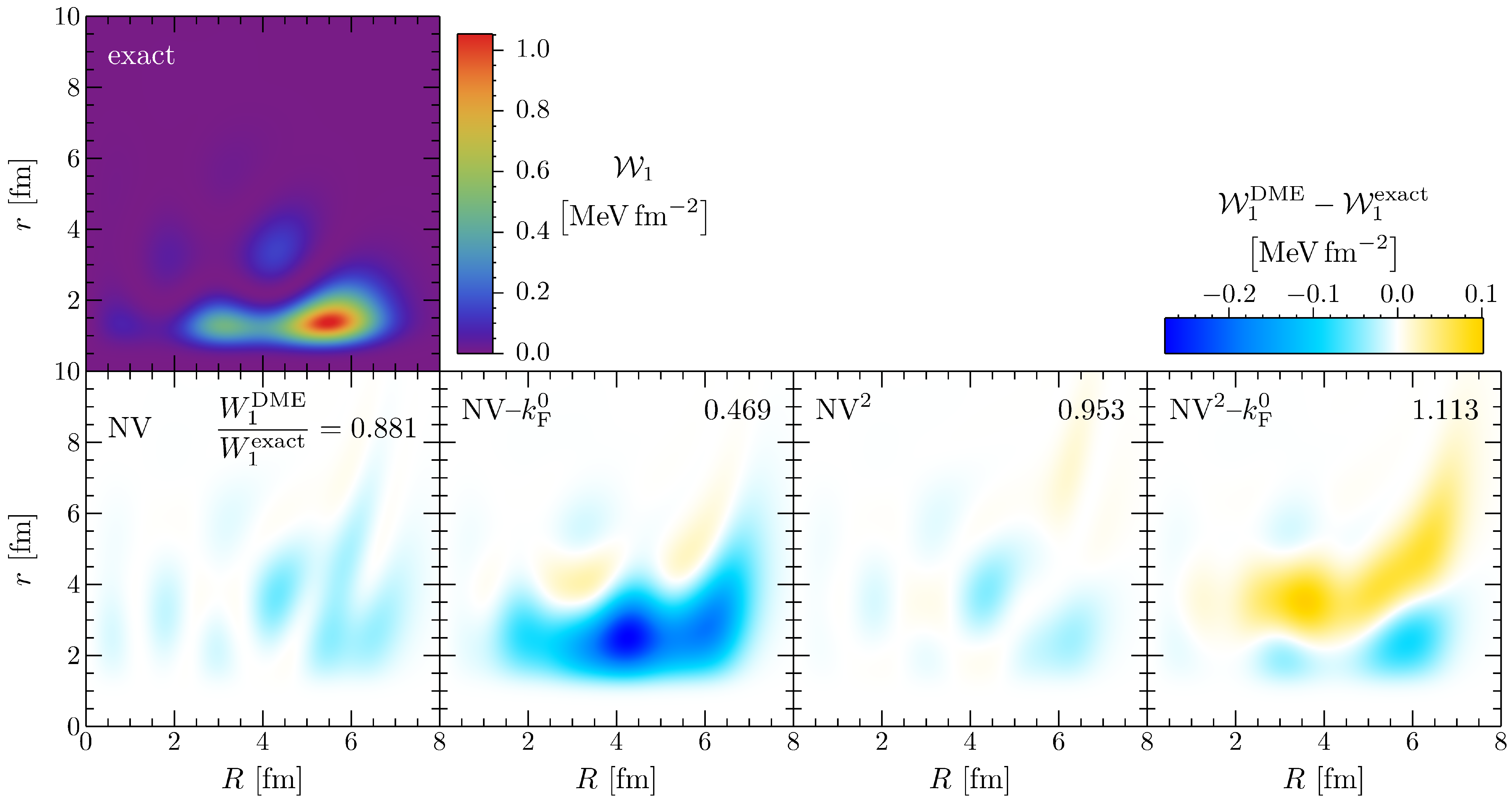}
\caption{\label{fig:1Integrand-Sn132}First panel: exact scalar-isovector exchange-energy integrand obtained for a regularized Yukawa interaction in $^{132}$Sn. Other panels: differences of DME approximations of this integrand and the exact integrand itself. Approximations obtained with isoscalar momentum scales for both neutron and proton density matrices are labeled with ``--$\kF^0$'' next to the abbreviated DME name. In every difference panel the value of the ratio of the DME-approximated energy and the exact energy is shown in the top right corner. The underlying orbitals are obtained from a self-consistent HF calculation with the SLy4-EDF.}
\end{figure*}

The right panel of \cref{fig:Ratios-np-SLy4} contains the ratios for the scalar-isovector energies $W_1$ as given by \cref{eq:IsovectorEnergy}. Here only the 6 asymmetric nuclei (with $N\neq Z$) in our set are considered since the isovector energies are completely negligible for the symmetric nuclei.
For most DMEs the ratios $W_1^\text{DME}/W_1^\text{exact}$ are further away from the ideal value of unity than in the isoscalar case. This can be understood when comparing the shape of the isoscalar part of the OBDM, which is a bulk quantity, to that of the isovector part, which is basically a neutron-excess density matrix. Thus, the region contributing the most to the isovector integral is located much closer to the nuclear surface where omitted higher-order corrections are expected to be more relevant. 
This is also clearly visible for $^{132}$Sn when comparing the scalar-isoscalar integrand $\mathcal{W}_0$ in the first panel of \cref{fig:0Integrand-Sn132} with the scalar-isovector integrand $\mathcal{W}_1$, which is defined as
\begin{align}
\label{eq:1Integrand}
    \mathcal{W}_1 (R,r) &\equiv \frac{3}{8} \int \! \dd\Omega_{\vR} \, \dd\Omega_{\vr} \, R^2 r^2 |\rho_1(\vR; \vr)|^2
    W_S^{\mathrm{LO}}(r) f(r)\,,
\end{align} 
and is depicted in the first panel of \cref{fig:1Integrand-Sn132}.
In addition, the energy contributions stem on average from a larger $r$ value in the isovector case for all considered nuclei, which again makes an accurate description harder when using DMEs. 

Nevertheless, the general trends are very similar for the isovector and the isoscalar energies. Notable exceptions are the BZ- and the Gaussian DME because their overestimations in the nuclear interior (also reflected in them not yielding the correct INM limit) matter less for the isovector part.

Overall, our results show that DMEs do not perform as well for scalar-isovector energies as for scalar-isoscalar energies, with typical accuracies being around 10\%.
For the considered asymmetric nuclei the magnitude of the scalar-isovector energies is on average only 1.3\% of the scalar-isoscalar contributions. Therefore, the worse accuracy in the isovector case has no relevant effect on the total energy reproduction, though it might be important when looking at nonbulk quantities.

\subsection{\label{sec:IsoscaMom}Isoscalar expansion momentum scale} 

So far, the results have been obtained by expanding neutron and proton density matrices separately as described in \cref{sec:DME} and subsequently forming the isoscalar and isovector parts by the appropriate sums, \cref{eq:IsoscalarSum,eq:IsovectorDiff}. However, this procedure yields EDFs where the terms that are normally isospin invariant (such as those proportional to $\rho_0^2$) also contain isospin-dependent parts, though their isospin symmetry is still conserved~\cite{Gebr10DME}.
Hence one might want to utilize another possibility that is to expand both the isoscalar and isovector parts as a whole. Then \cref{eq:DME2ndOrderAngAv} becomes
\begin{align}
 \rho_t(\vR; \vr) & \approx \Pi_0(k \rNN)\rho_t(\vRNN)
 +\frac{\Pi_2(k \rNN)}{6} r^2
  \nonumber \\  & \quad \null \times
  \left[\frac{1}{4} \vnabla^2 \rho_t(\vRNN) - \tau_{t}(\vRNN) + \frac{3}{5} k^2 \rho_t(\vRNN)\right]\,,
\end{align}
where $t=0,1$.
Using different momentum scales for the isoscalar and isovector expansions leads to additional complications. Therefore, we follow Ref.~\cite{Doba10DMEGogny}
and simply use the isoscalar variants of \cref{eq:FermiMom,eq:CampiMom},
\begin{align}
    \kF^0(\vR) &\equiv \left[ \frac{3\pi^2}{2} \rho_0(\vR) \right]^{1/3} \,, \\
    \kCB^0(\vR) &\equiv \left\{ \frac{5}{3\rho_0(\vR)} \left[\tau_0(\vR) - \frac{1}{4} \vnabla^2 \rho_0(\vR) \right] \right\}^{1/2} \,,
\end{align}
for all OBDM parts. Then, this prescription is equivalent to using \cref{eq:DME2ndOrderAngAv} but with the same momentum scale for both neutrons and protons. 
Because $k_\mathrm{F(C)}^\mathrm{n}(\vR) \approx k_\mathrm{F(C)}^0(\vR) \approx k_\mathrm{F(C)}^\mathrm{p}(\vR)$ one may expect the results to not be significantly different for either of the momentum scales.
But in the particular case of pure isovector quantities using $\kF^0(\vR)$ or $\kFC^0(\vR)$ could be much worse as this effectively results in approximating the difference of neutron and proton density matrices with a momentum scale that assumes their similarity. 
This can also be viewed as employing a single-species procedure to approximate the neutron-skin density matrix, which almost never behaves like a single-species density matrix.

This is confirmed by the panels in the second row of \cref{fig:1Integrand-Sn132}, which display the differences between DME-approximated and exact scalar-isovector integrands $\mathcal{W}_1$. We show them for the \mbox{NV-} and NV$^2$-DMEs, both for separate neutron/proton momentum scales and for the isovector momentum scale $\kF^0$. The expected much larger (local) deviations in the latter case are clearly visible. This is similar for the other DMEs that are not displayed and translates also to the energy ratios $W_1^\text{DME}/W_1^\text{exact}$.

\begin{figure}[t]
\includegraphics[width=1\linewidth]{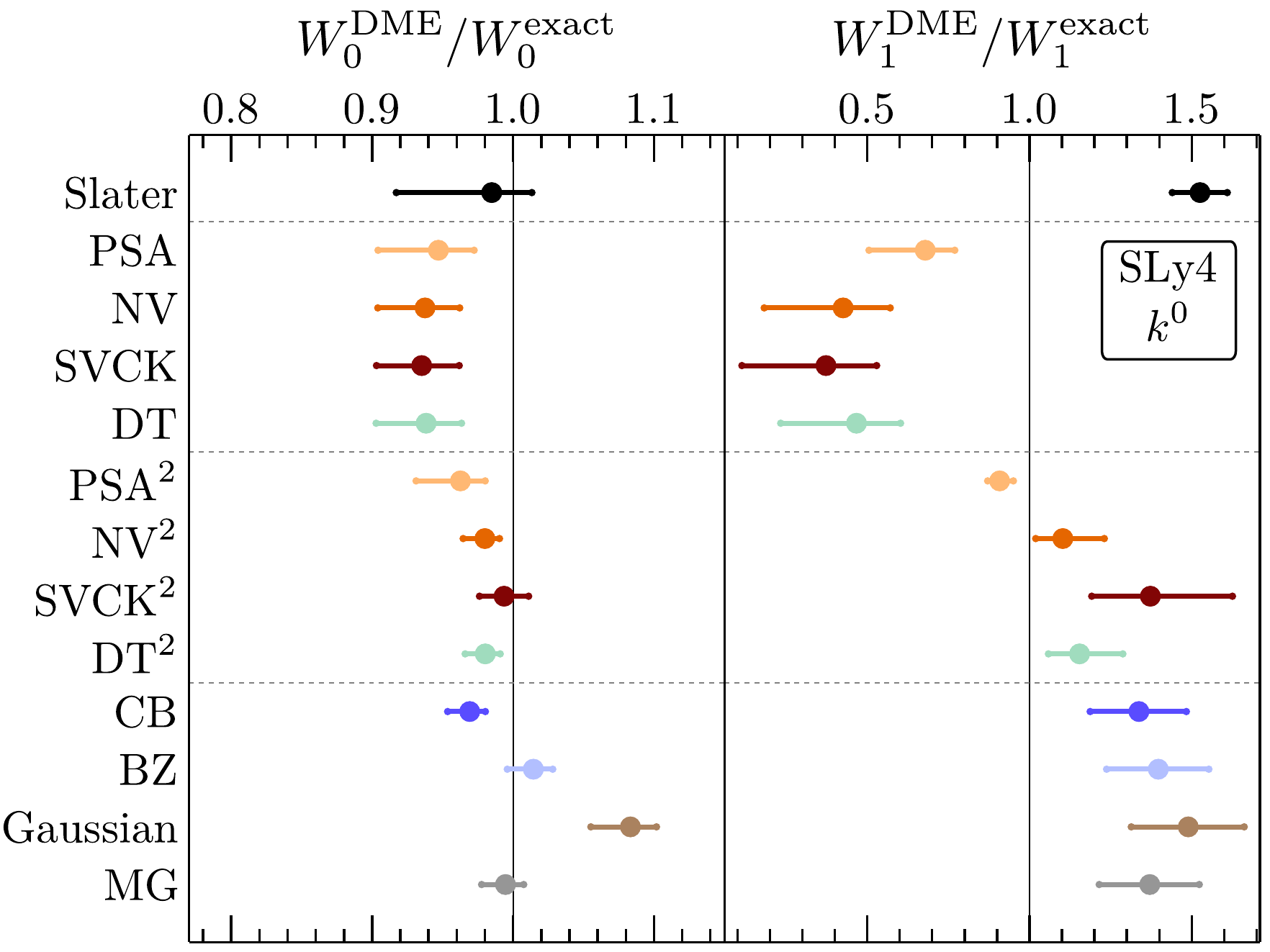}
\caption{\label{fig:Ratios-01-SLy4}Same as \cref{fig:Ratios-np-SLy4} without the ratios for individual nuclei. Unlike in \cref{fig:Ratios-np-SLy4}, here both panels show expansions with isoscalar momentum scales. Note the different axis scale of this figure when comparing to other figures.}
\end{figure}

In the right panel of \cref{fig:Ratios-01-SLy4} we show these ratios, but unlike in \cref{fig:Ratios-np-SLy4} here the values are obtained by using the isoscalar variants of the momentum scales. 
The results are much worse for the isoscalar momentum scale: the average ratios range from 0.37 to 1.53 and are in all cases further away from unity than with separate momentum scales. 
However, the scalar-isoscalar energies are almost identical for isoscalar (\cref{fig:Ratios-01-SLy4}) and separate momentum scales for the two species (\cref{fig:Ratios-np-SLy4}). As explained, both observations are expected.

Whether one deems using DMEs with isoscalar momentum scales acceptable or not in light of these findings depends very much on the case at hand. The poor accuracy of the very small scalar-isovector energies effectively does not matter when one is only interested in a good description of the total energies,\footnote{For some DMEs the total energy reproduction is even slightly better with the isoscalar momentum scale due to cancellations of errors between scalar-isoscalar and scalar-isovector energies.} but again this might not be true for isovector and differential quantities, such as differences along isotope chains.

\subsection{\label{sec:DepOrbitals}Dependence on orbitals} 

In this subsection we want to answer the question whether the results reported above depend sensitively on details of the orbitals. The orbitals used so far were obtained from self-consistent HF calculations with the SLy4-EDF. We now switch to orbitals from a simple isotropic harmonic oscillator (HO) with frequency  $\hbar\omega = 10 \unit{MeV}$. As can be seen in \cref{fig:DensityDistributions} they are quite well suited to provide a less realistic counterpart to the SLy4 orbitals. We consider the same 11 (6) nuclei as before for the scalar-isoscalar (scalar-isovector) energies. 

\begin{figure}[t]
\includegraphics[width=1\linewidth]{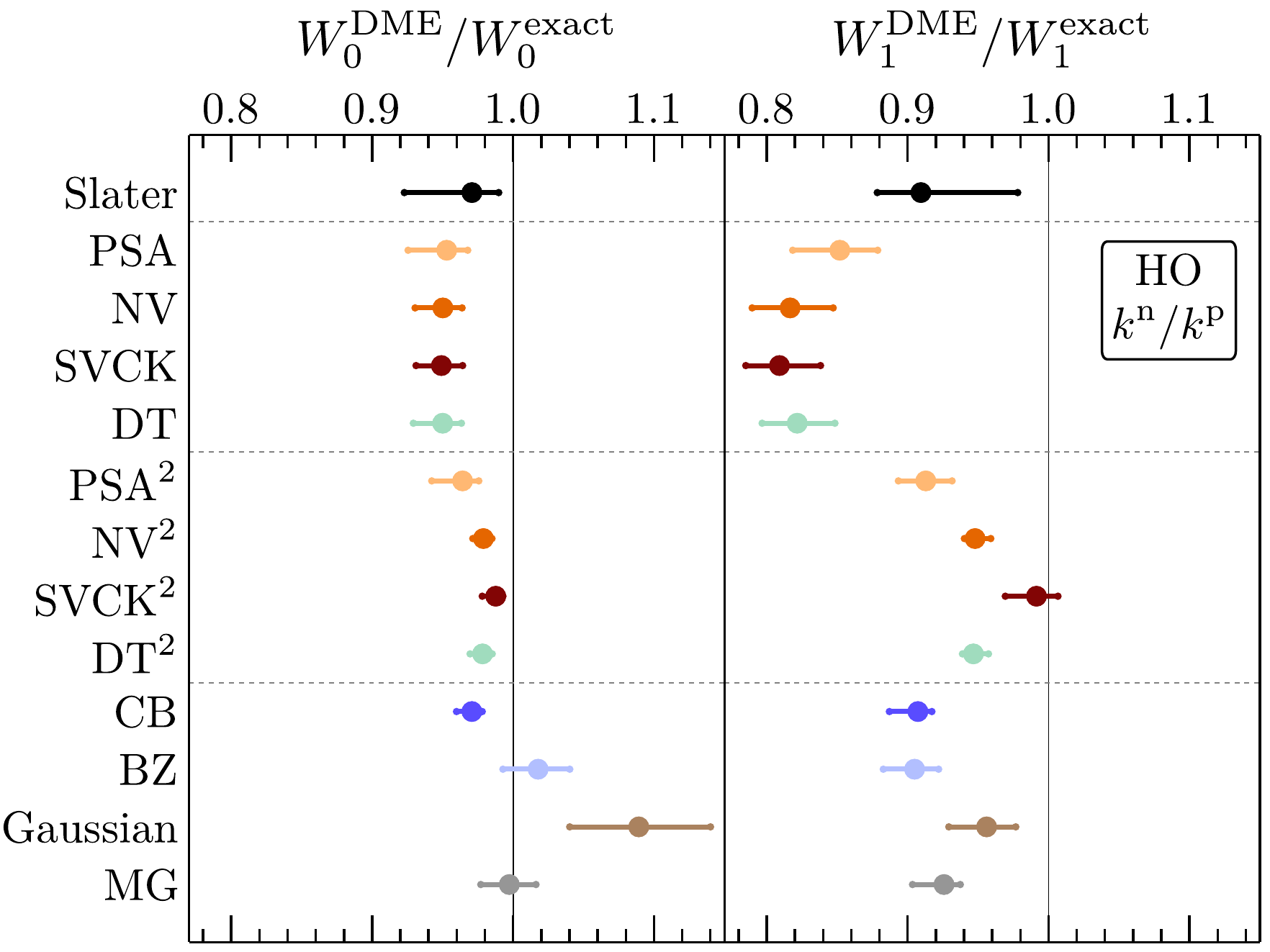}
\caption{\label{fig:Ratios-np-HO}Same as \cref{fig:Ratios-np-SLy4} without the ratios for individual nuclei. The nuclei are given in terms of orbitals from an isotropic HO with $\hbar\omega = 10\unit{MeV}.$ In both panels the DMEs are used with separate momentum scales for neutrons and protons.}
\end{figure}

Changing back to expansions with separate momentum scales for neutrons and protons we show the ratios $W_0^\text{DME}/W_0^\text{exact}$ and $W_1^\text{DME}/W_1^\text{exact}$ in \cref{fig:Ratios-np-HO}.
For both scalar-isoscalar and scalar-isovector energies the results are very similar to the SLy4 results given in \cref{fig:Ratios-np-SLy4}.
The main difference is that the spread between the smallest and the largest ratios is typically slightly smaller in the case of HO orbitals but the ranking of the DME variants according to the accuracy of their Yukawa exchange energy reproduction is very similar.

As an additional check we use orbitals obtained from spherical HF calculations employing the 1.8/2.0 (EM) interaction~\cite{Hebe11fits}. 
For selected nuclei, the corresponding isoscalar density distributions are shown in \cref{fig:DensityDistributions}. 
Again, the accuracy ranking of the DMEs is virtually the same. Detailed results for these orbitals are presented in the Supplemental Material~\cite{Supplemental}.

Overall the investigations of this section strongly indicate that previous findings regarding the accuracy of reproducing Yukawa exchange energies are generally true, i.e., do not sensitively depend on orbitals.

\subsection{\label{sec:DepInteraction}Dependence on interaction} 

Exchange energies from interactions with shorter ranges are expected to be better reproduced by DMEs.
Our tests confirm such behavior. In particular, DMEs are exact in the limit of vanishing interaction range. 
But what about the opposite limit? Consider
    \begin{equation}
    W_{S}^{\mathrm{LO}}(m, r) \equiv  \frac{m_\pi^3}{12\pi} \left( \frac{g_{A}}{2 F_\pi} \right)^2 \frac{\ee^{-m r}}{m_\pi r} \,,
    \end{equation}
where the parameter $m$ is the reciprocal of the interaction range. One-pion exchange is obtained for $m=m_\pi$ and the infinite-range limit (i.e., the Coulomb interaction) for $m=0$. 

\begin{figure}[t]
  \includegraphics[width=\linewidth]{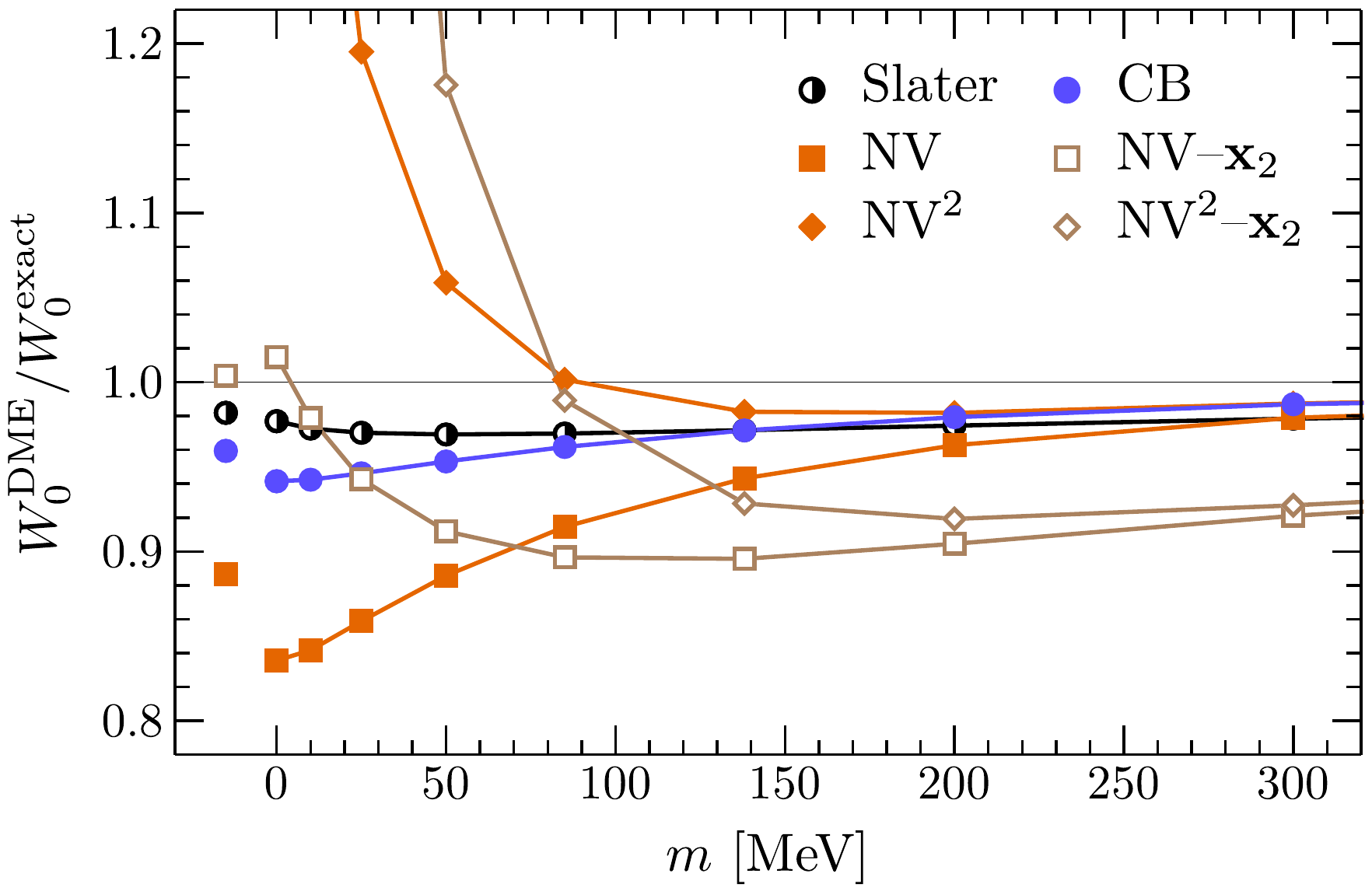}
\caption{\label{fig:InteractionRange}Ratio of DME-approximated and exact scalar-isosclar exchange-energy contribution for a regularized Yukawa interaction. The parameter $m$ corresponds to the reciprocal of the interaction range. The values are averages over 11 closed-shell nuclei obtained from a self-consistent HF calculation with the SLy4-EDF. Density-matrix expansions are used with separate momentum scales for neutrons and protons and the expansions are about the two-particle center of mass $\vRNN$ except for the cases marked with ``--$\vx_2$'' where instead they are about the position of one particle. Extra points on the left show the result at $m=0$ without regulators.}
\end{figure}

In \cref{fig:InteractionRange} we plot the scalar-isoscalar energy ratios $W_0^\text{DME}/W_0^\text{exact}$ for this interaction as a function of $m$, where again each point is averaged over the same 11 nuclei obtained from SLy4-EDF orbitals as in \cref{sec:Isosca,sec:Isovec,sec:IsoscaMom}. The interaction is also regularized as before [see \cref{eq:RegReplace,eq:Regulator}]. The energy ratios are shown for $m=0,$ 10, 25, 50, 85, 138.03, 200, and $300\unit{MeV}$. In addition, for each DME a single additional point, which corresponds to the value at $m=0$ without regulators, is drawn on the very left.
\Cref{fig:InteractionRange} contains the results for \mbox{Slater-,} \mbox{NV-,} \mbox{NV$^2$-,} and CB-DMEs. The behavior for the other second-order DMEs with $\kF$ ($\kFC$) is similar to the NV/NV$^2$ (CB) trends.

For large interaction ranges the DME exchange-energy integrals, \cref{eq:IsoscalarEnergy,eq:IsovectorEnergy}, have to be carried out up to very high $r$ values to obtain converged results. This is especially important for full-square DMEs because their oscillations with significant amplitudes occur for particularly large $r$ in regions of small expansion momenta. For one-pion exchange ($m = 138.03\unit{MeV}$) these regions are damped, but when the interaction falls off much more slowly they contribute nonnegligibly. Thus, we calculate the integrals for $m \leq 25\unit{MeV}$ analytically without the regulator by employing a strategy proposed in \rcite{Fabr12BesInt} and add to that the correction from the regulator, which can easily be calculated numerically due to its short range. Details on this procedure and the relevant analytical expressions are provided in the Supplemental Material~\cite{Supplemental}. Note that when using an isoscalar momentum scale those analytical integrals can also be obtained with the Mathematica package of \rcite{Gebr100BesInt}.

As expected, the NV-DME results significantly deteriorate with increasing interaction range (i.e., decreasing $m$) and even more so do the NV$^2$-DME results. This is in agreement with results from \rcite{Camp78DME}. The worse accuracy for the full-square variant is due to the unwanted large-$r$ bump of the expansion (see \cref{fig:rho2} for an example) getting probed more for larger interaction ranges. 

We consider here also DMEs around $\vx_2$ as discussed in \cref{sec:coords}, which have been reported to yield good results with the Coulomb interaction in molecular systems~\cite{Koeh96MolecDME}. These are labeled with an additional ``--$\vx_2$'' in \cref{fig:InteractionRange}. While they perform worse for small ranges, the NV-DME about $\vx_2$ produces much better results for large interaction ranges than its conventional counterpart. This is despite the angle averaging being performed before squaring the density matrix. The opposite order of these two operations yields a different expression for expansions about $\vx_2$ and should also be investigated in the future.
The improved energy reproduction also holds when neglecting the regulator and agrees qualitatively with the molecular-physics result of \rcite{Koeh96MolecDME} despite the different considered systems. This confirms the conclusion of the previous section.

As elaborated on in \cref{sec:coords}, generalizing DMEs that use $\kFC$ to expansions about $\vx_2$ leads to complications and hence we give the CB-DME results only about $\vR$. We observe that the CB-DME accuracy is significantly less range dependent than the second-order $\kF$-DMEs. 
The performance of the Slater approximation (which is the same for expansions about $\vR$ and $\vx_2$) is even less range dependent. 

We also note that for PSA- and PSA$^2$-DME the energies are infinite in the Coulomb limit, independent of the expansion point, due to insufficient convergence of these DME variants, see also the corresponding remark in \cref{sec:DMEvar}. 
Depending on the asymptotic behavior of the orbitals, the Coulomb exchange integrals can diverge for any full-square DME, see \rcite{Camp78DME} for an example.

The Yukawa interaction considered in the previous subsections contains another length scale in addition to $m$: the regulator cutoff $R_0$, see \cref{eq:Regulator}. The dependence of the DME accuracy on $R_0$ is straightforward. As the regulator cuts off only short-distance parts of the interaction where DMEs work very well, larger regulator cutoffs correspond to a worse overall reproduction of exact exchange energies and a larger spread of the accuracies for different systems. Detailed values for different regulator cutoffs as well as for the EKM regulator~\cite{Epel15improved}
\begin{equation}
\label{eq:EKMRegulator}
    f(r) = \left[ 1 - \exp\left( -\frac{r^2}{R_0^2} \right) \right]^6
\end{equation}
are provided in the Supplemental Material~\cite{Supplemental}.

There we also show results for the finite-range parts of the Gogny D1S interaction~\cite{Berg91GognyD1S}, a successful phenomenological pseudopotential that was also considered in DME studies of Refs.~\cite{Doba10DMEGogny, Carl10DMEConv}. Its finite-range parts are two Gaussians that contribute with different signs to the exchange energy. The resulting cancellations change the ranking of the DMEs according to their accuracy in minor details but the overall conclusions of this work are still valid.

Summing up, for not-too-long-range NN forces such as one-pion exchange, DMEs around the center of mass $\vR$ yield the best results. This is not the case for the Coulomb interaction where for instance in the case of the NV-DME expanding about $\vx_2$ is superior.

\section{Summary and outlook}\label{sec:summary}

Empirical EDFs have been broadly successful in describing nuclear properties across the table of nuclides. Recently, EDFs have included long-range pion contributions from chiral EFT with encouraging and puzzling results~\cite{Nava18DMEEDF}. In this work, we therefore have taken the first steps with a detailed re-examination of the DME implementation. 
To this end, we compared several zeroth- and second-order DMEs for scalar parts of OBDMs, focusing on the accurate non-self-consistent reproduction of exact Yukawa exchange energies in closed-shell nuclei. 

In general, all considered DMEs approximate the investigated exchange energies very well.
Of those DMEs that do not lead to more than two derivatives in any EDF term (like conventional Skyrme EDFs) we find best energy reproduction for the Slater approximation, although locally it approximates the energy integrands worse than second-order DMEs. 
When allowing for EDF terms with more than two derivatives, but still using only the standard Skyrme densities, one can also employ the full-square DMEs.
These perform better than their truncated-square counterparts and the ones that use $\kFC$ as their momentum scale. 
Overall we find best results for the SVCK$^2$- and MG-DMEs, although the latter yields the wrong INM limit.

Regarding a good reproduction of scalar-isovector energies we find that it is crucial to treat neutrons and protons separately in DMEs. Using a single isoscalar momentum scale can lead to results wrong by more than 50\%, though the effect on the total exchange energy is very small due to the small absolute size of isovector contributions. 

All these findings are robust in the sense that they hold along the entire nuclear mass range, are confirmed also for less realistic orbital shapes, and are valid for different regulators and interaction ranges (except for very-long-range interactions, see \cref{sec:DepInteraction}). 

Our results put the DME applicability for long-range pion contributions on a solid footing, showing that the different DME choices generally lead to tolerable variations. This does therefore not resolve the puzzles in the EDF performances with chiral physics included.

All results in this paper are based on non-self-consistent tests and should therefore be regarded as provisional. For instance, it is at this stage unclear how local errors in the reproduction of exchange-energy integrands (e.g., see Slater approximation in \cref{fig:0Integrand-Sn132}) influence the results of the self-consistency loop in an EDF calculation of nuclei.
Hence, one of next steps is to implement the findings of this work into EDFs like the ones of \rcite{Nava18DMEEDF}. This could also bring us closer to answering if explicit pions are needed for higher EDF accuracies.
However, we believe the present findings suggest the EDF improvement coming from an enhanced DME treatment will be minor, especially considering that the Skyrme couplings get refitted after incorporating the DME in the approach of \rcite{Nava18DMEEDF}.
In general, EDF practitioners can test the performance of the DME variant of their choice by switching to one of the other variants discussed in this work.
If the results are quantitatively very similar, this suggests that further EDF improvements need to come from elsewhere, and not from DME improvements.

This paper only dealt with the application of DMEs to NN forces. In the 3N sector, even more choices have to be made regarding the DME and the authors of \rcite{Nava18DMEEDF} report that the inclusion of 3N forces degrades the quality of their EDFs at every considered chiral order. 
In addition, we did not consider vector parts of the OBDM, or
DME terms with an odd number of derivatives that are relevant in not-time-reversal-invariant systems.
These topics are left for future investigations.

\begin{acknowledgments}

We thank Jacek Dobaczewski for helpful discussions, Jan Hoppe for providing the 1.8/2.0 (EM) HF orbitals, and Matthias Heinz for comments on the manuscript.
The work of L.Z., E.A.C.P., and A.S. was supported in part by the Deutsche Forschungsgemeinschaft (DFG, German Research Foundation) -- {Project-ID} 279384907 -- SFB 1245 and by the BMBF Contract No.~05P18RDFN1.
The work of E.A.C.P. was also performed in part under the auspices of the U.S. Department of Energy by Lawrence Livermore National Laboratory under Contract No. DE-AC52-07NA27344.
The work of R.J.F. was supported in part by the National Science Foundation under Grant No.~PHY--1913069 and by the NUCLEI SciDAC Collaboration under Department of Energy MSU Subcontract RC107839-OSU. 
The work of S.K.B. was supported in part by the National Science Foundation under Grants No.~PHY--1713901 and No.~PHY--2013047, and by the NUCLEI SciDAC Collaboration under Department of Energy Grant No.~de-sc0018083.
\end{acknowledgments}

\bibliography{literature}

\clearpage

\section*{Supplemental material}

\subsection*{Hartree-Fock orbitals for 1.8/2.0 (EM) interaction}

\begin{figure}[h]
\includegraphics[width=1\linewidth]{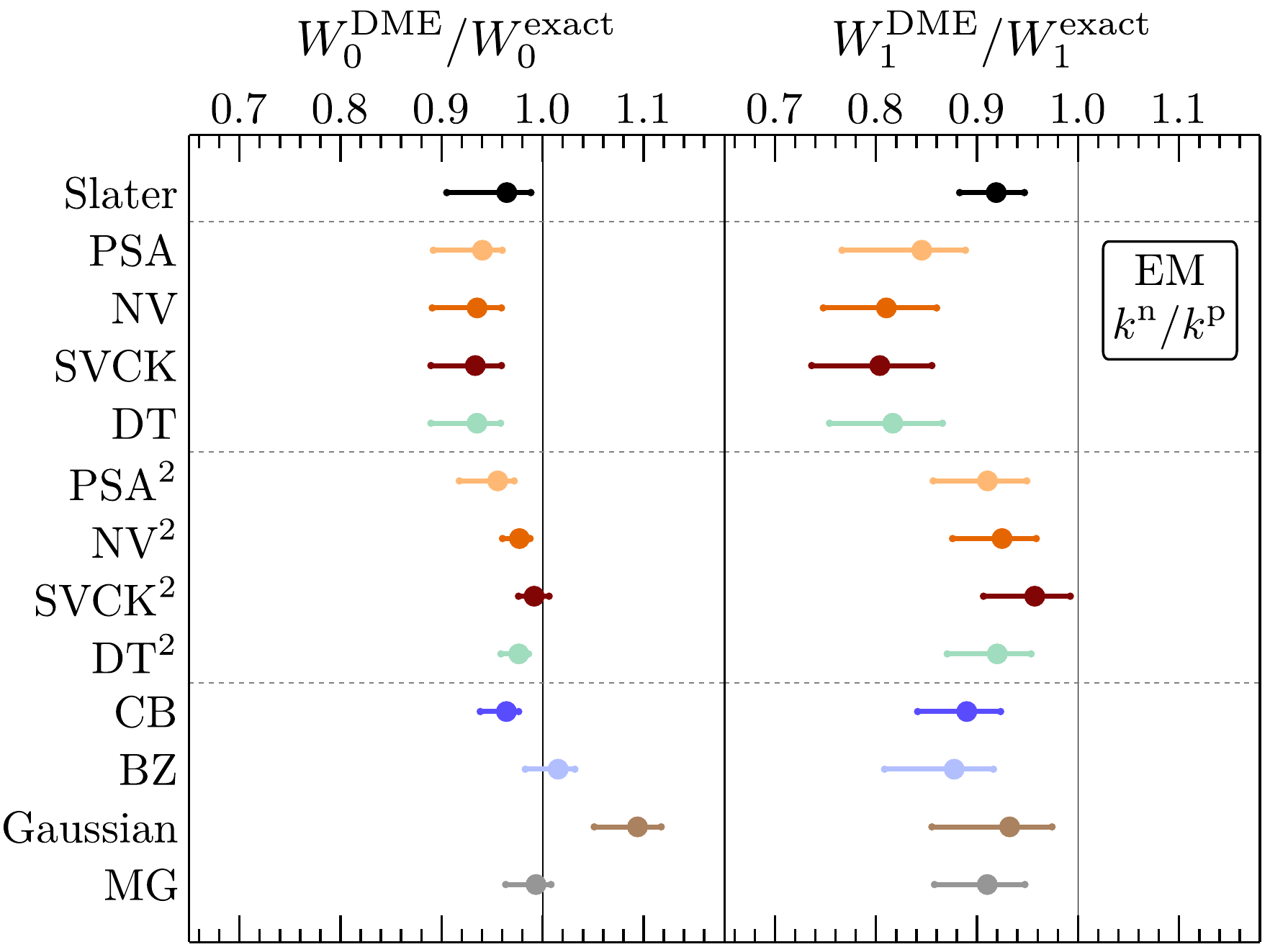}
\caption{\label{fig:Ratios-01-EM}Ratios of DME-approximated and exact exchange-energy contributions for a regularized Yukawa interaction. 
For every DME variant the average over a set of nuclei is shown together with a bar ranging from the smallest to the largest ratio observed. 
The underlying orbitals are obtained from a self-consistent HF calculation with the 1.8/2.0 (EM) interaction. The left panel shows the results for the scalar-isoscalar contributions, where 11 closed-shell nuclei are considered (see text); the right panel shows the results for the scalar-isovector contributions (from 6 closed-shell nuclei). In both panels the DMEs are used with separate momentum scales for neutrons and protons.
}
\end{figure}

In addition to the SLy4 orbitals, we show results for orbitals from a spherical HF calculation based on a chiral low-momentum two- plus three-nucleon interaction~\cite{Hebe11fits}, 1.8/2.0 (EM), which has been used widely in {\it ab initio} calculations of medium-mass nuclei. The HF orbitals are expanded in an HO basis with $\hbar\omega = 16\unit{MeV}$ and $e_{\text{max}} \leq 12$, and the three-body configurations are included up to $E_{3\text{max}} \leq 16$.
We consider the same nuclei as in the main text.
\Cref{fig:Ratios-01-EM} shows that the DME performance is very similar for these orbitals, with slightly larger spread and slightly worse energy reproduction. This further supports the general conclusions of our paper.

\ \\ \ \\ \ \\ \ \\ \ \\ \ \\
\subsection*{Different regulators and cutoff choices}

\begin{figure}[h]
\includegraphics[width=1\linewidth]{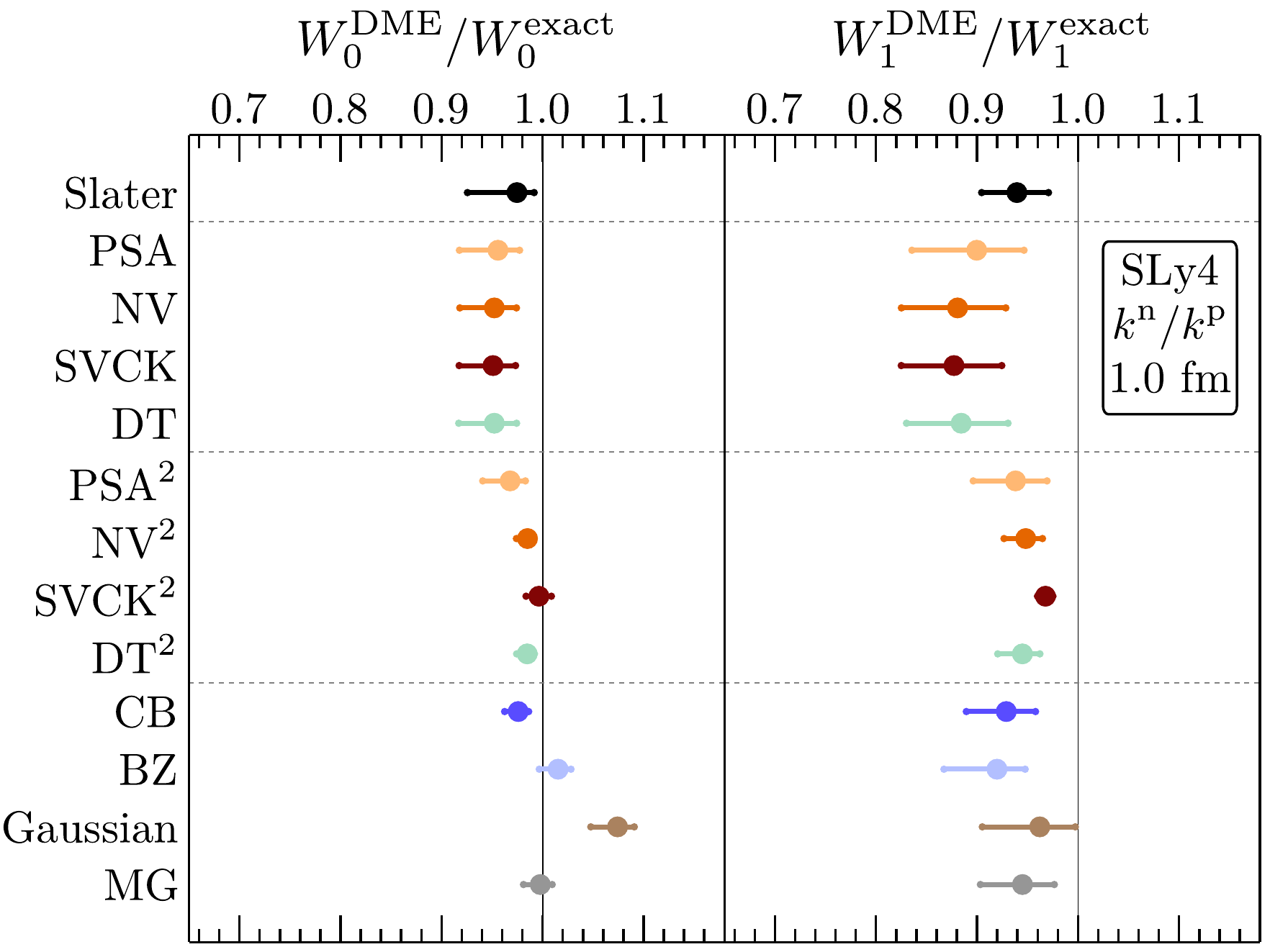}
\caption{\label{fig:R0_10}Same as \cref{fig:Ratios-01-EM} but with orbitals from a self-consistent HF calculation with the SLy4-EDF and for a regulator cutoff $R_0 = 1.0\unit{fm}$.
}
\end{figure}

\begin{figure}[b]
\includegraphics[width=1\linewidth]{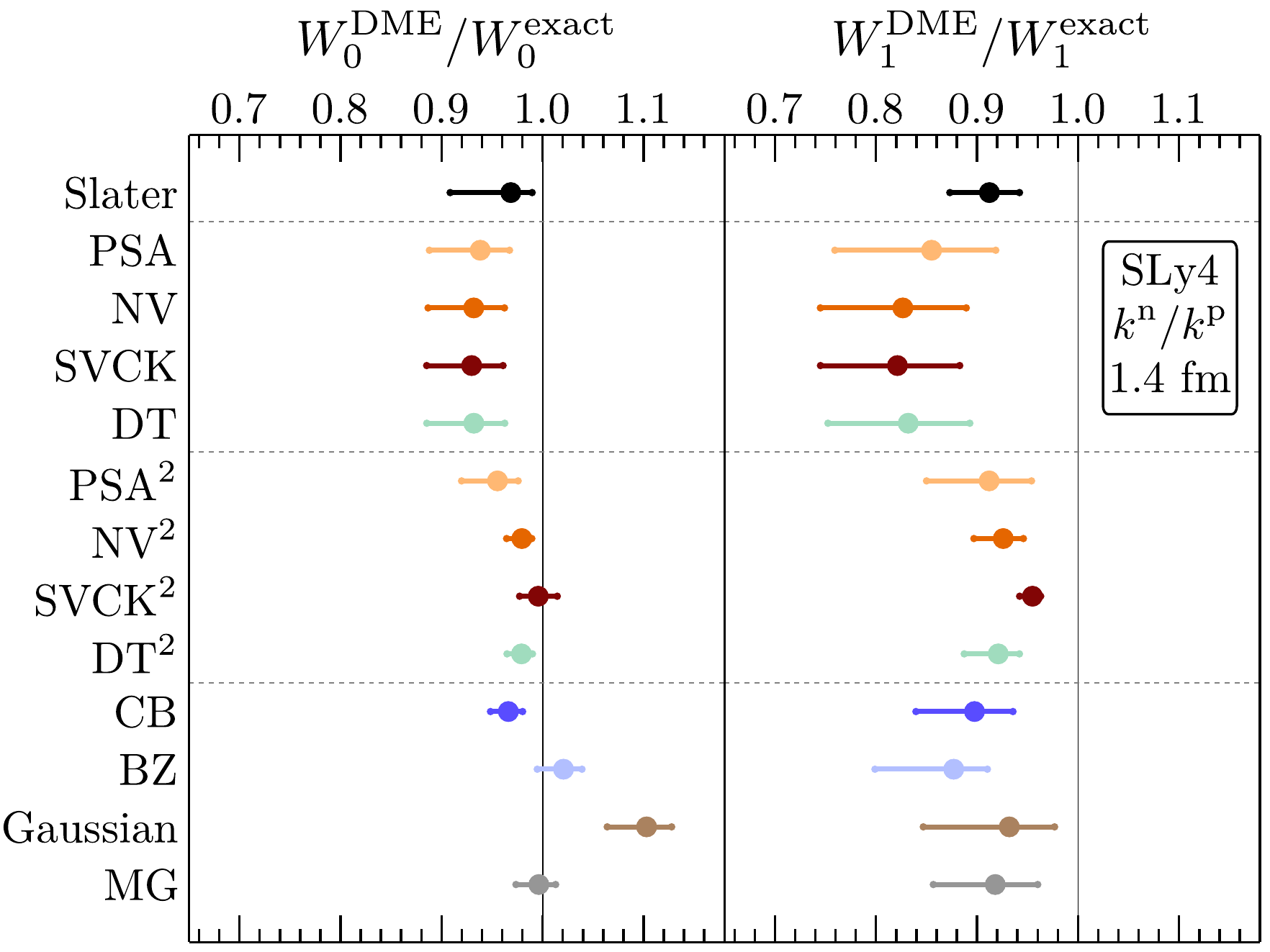}
\caption{\label{fig:R0_14}Same as \cref{fig:R0_10} but for a cutoff $R_0 = 1.4\unit{fm}$.}
\end{figure}

\begin{figure}[t]
\includegraphics[width=1\linewidth]{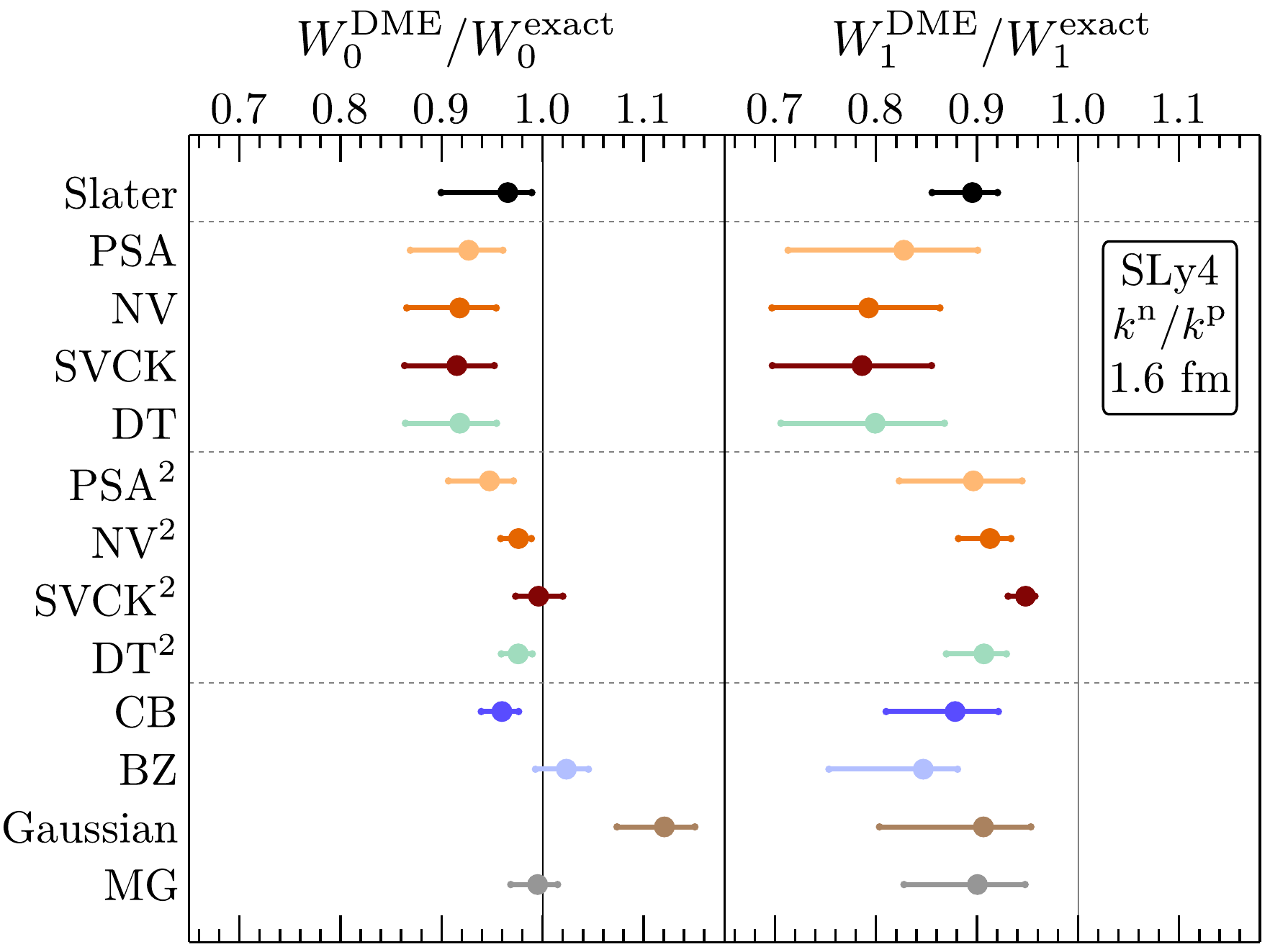}
\caption{\label{fig:R0_16}Same as \cref{fig:R0_10} but for a cutoff $R_0 = 1.6\unit{fm}$.}
\end{figure}

\begin{figure}[t]
\includegraphics[width=1\linewidth]{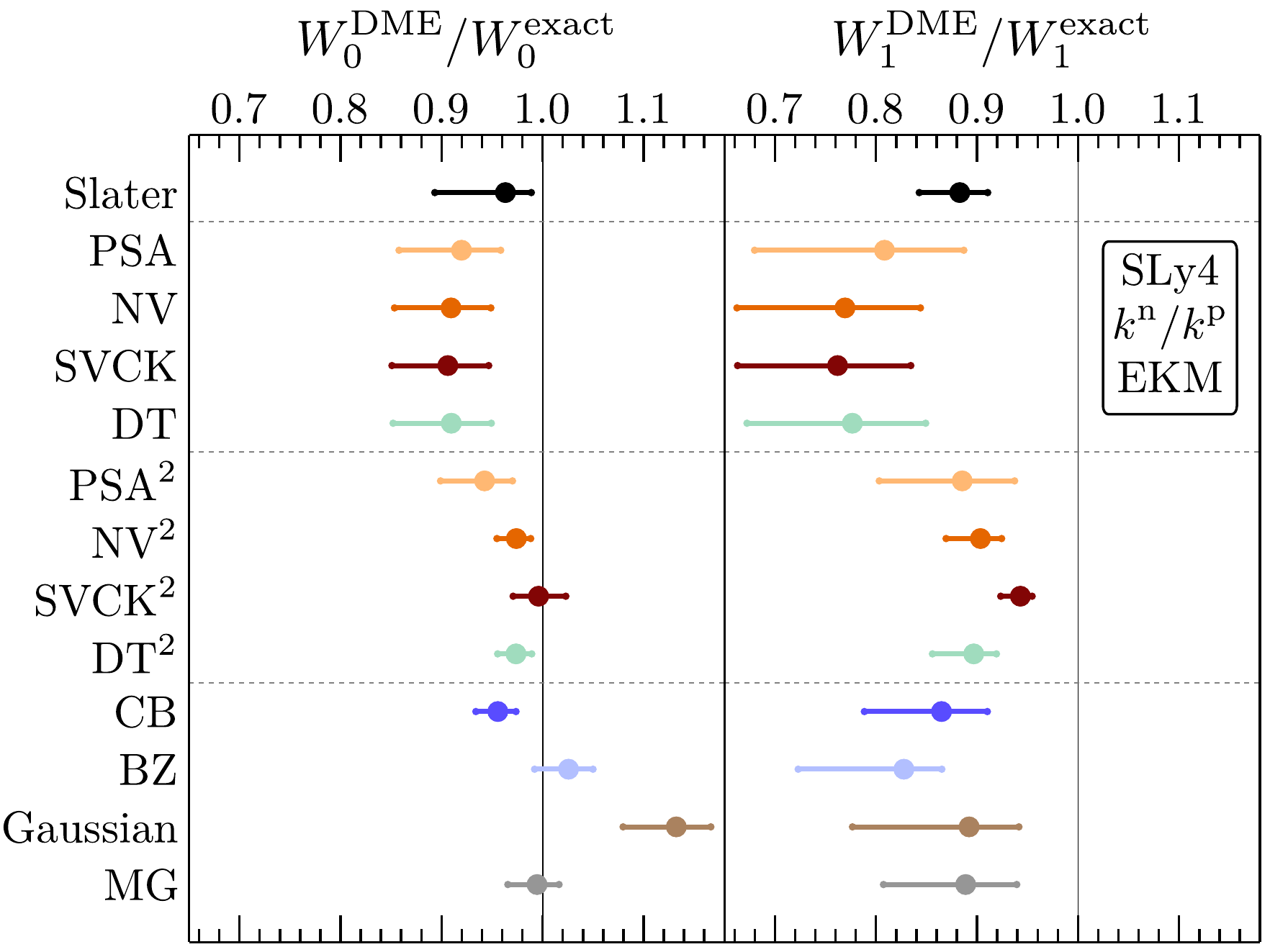}
\caption{\label{fig:EKM}Same as \cref{fig:R0_10} but for the EKM regulator, given in \cref{eq:EKMRegulator}, with cutoff $R_0 = 1.0\unit{fm}$.}
\end{figure}

In addition, we explore different regulators and cutoff choices for the Yukawa interaction. In \cref{fig:R0_10,fig:R0_14,fig:R0_16,fig:EKM}, we show results for the same local regulator with different cutoffs $R_0 = 1.0\unit{fm}$, 
$1.4\unit{fm}$, and $1.6\unit{fm}$, as well as for the EKM regulator, \cref{eq:EKMRegulator}, with cutoff $R_0 = 1.0\unit{fm}$~\cite{Epel15improved}.
The figures show that larger cutoffs correspond to a worse overall reproduction of exact exchange energies and a larger spread of the accuracies for different systems. These observations agree with expectations as the regulators cut off only short-distance parts of the interaction so that only the long-distance parts, where DMEs do not work as well, contribute to the energy.

\FloatBarrier
\subsection*{Application to a Gogny interaction}

\begin{figure}[t]
\includegraphics[width=1\linewidth]{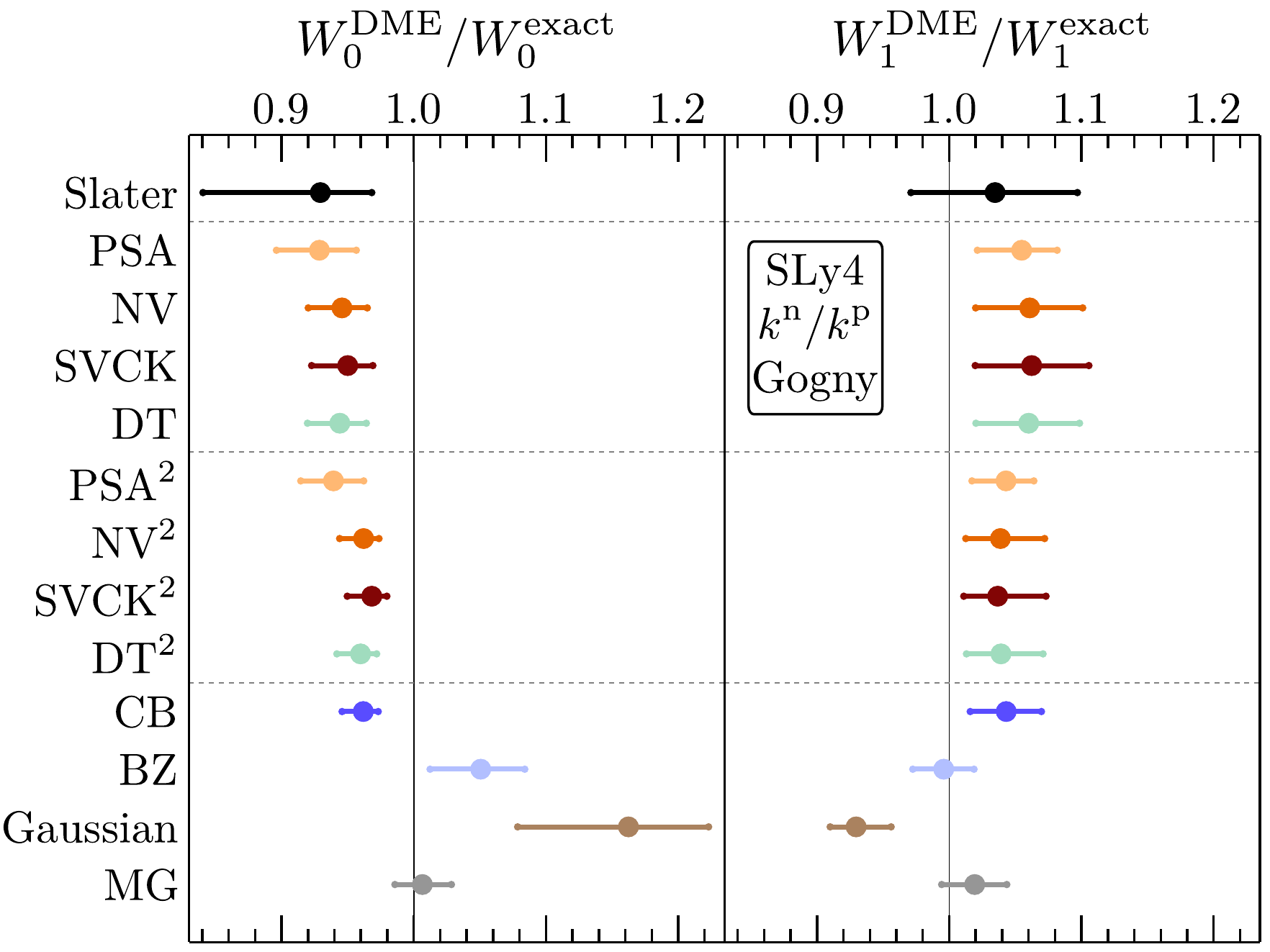}
\caption{\label{fig:GognyNP}Same as \cref{fig:R0_10} but for the finite-range parts of the Gogny D1S interaction.
}
\end{figure}

\begin{figure}[t]
\includegraphics[width=1\linewidth]{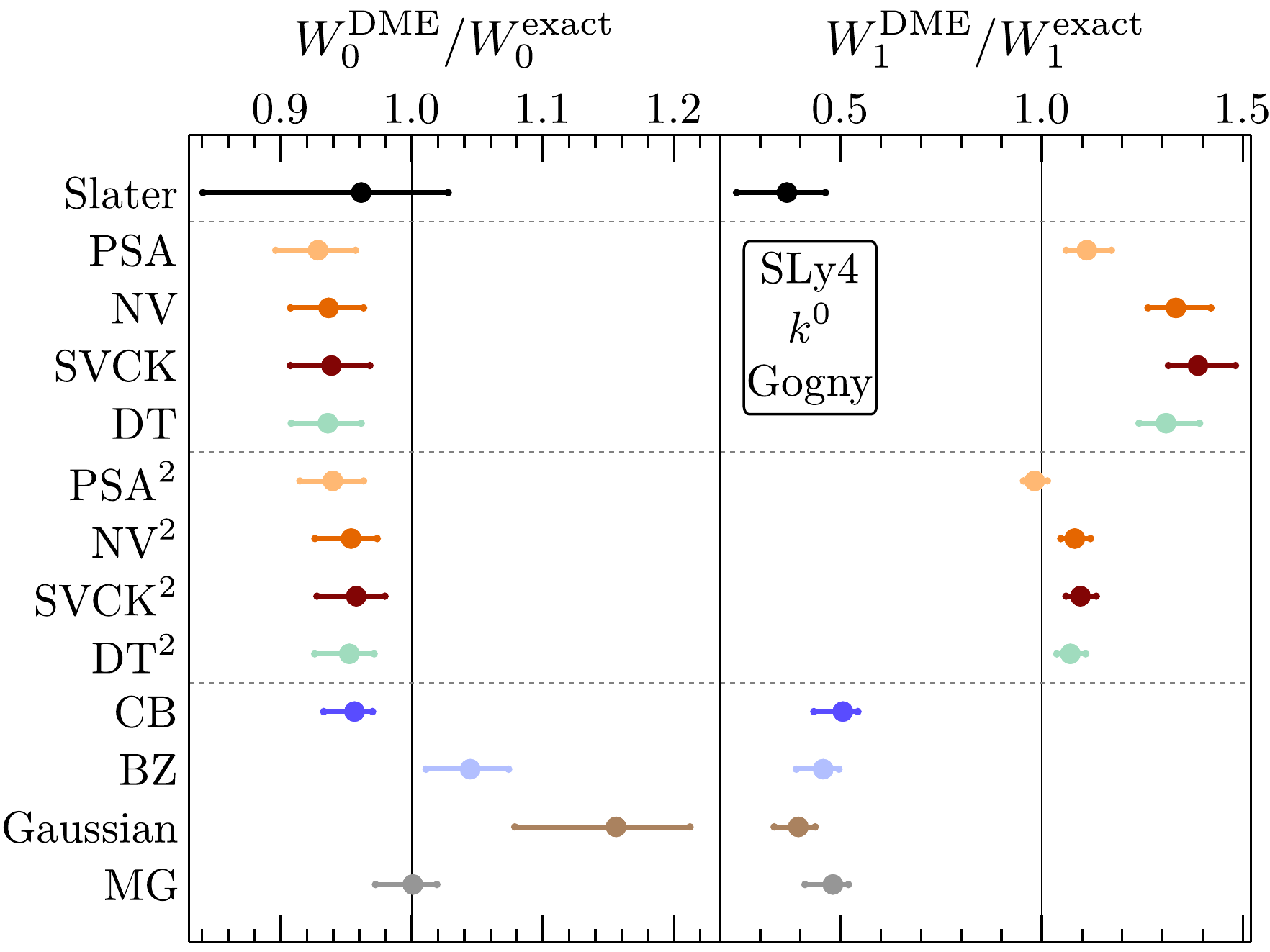}
\caption{\label{fig:Gogny01}Same as \cref{fig:GognyNP} but for expansions with isoscalar momentum scales.}
\end{figure}

As a final test, we explore the DME performances for the finite-range parts of the Gogny D1S interaction~\cite{Berg91GognyD1S}. 
The finite-range parts are given by a sum of two Gaussians which contribute with different signs to the exchange energy.
The results are shown in \cref{fig:GognyNP} for expansions with individual momentum scales for neutrons and protons and in \cref{fig:Gogny01} for expansions using isoscalar momentum scales for both species.
As for the Yukawa interaction the scalar-isovector energy reproduction is much worse when using the isoscalar momentum scale (except for the PSA$^2$-DME). 
One difference to the Yukawa-interaction results lies in the improvement from using full squares rather than truncated squares, which is smaller here. This is because the additional term in the full square affects mainly the large-$r$ behavior, which is not much probed by the Gaussians.
In addition, the ratios obtained for the Slater approximation depend significantly more on the nucleus in the present case.
This indicates that the cancellations of large local under- and overestimations as present in this approximation (see, e.g., \cref{fig:0Integrand-Sn132}) can be quite sensitive to system details.
Note that the DME accuracies for every individual Gaussian in the D1S interaction are much better than the total accuracies shown in \cref{fig:GognyNP,fig:Gogny01} because of cancellations between the two terms.
The overall conclusions of our paper are supported by these results.

\ \\
\ \\

\clearpage
\onecolumngrid

\subsection*{Semi-analytical EDF expressions}

To calculate the scalar-isoscalar DME exchange-energy contribution for regularized Yukawa interactions with very long ranges $m^{-1}$ we split the energy into two parts, 
\begin{equation}
    W_0 = W_0^\infty - W_0^\text{reg} \,,
\end{equation}
where 
\begin{equation}
    W_0^\infty \equiv -\frac{9}{8} \int \! \dd\vR\, \dd\vr \,  |\rho_0(\vR; \vr)|^2 W_S^{\mathrm{LO}}(m,r)
\end{equation} 
is the exchange-energy contribution without regulators and 
\begin{equation}
    W_0^\text{reg} \equiv -\frac{9}{8} \int \! \dd\vR\, \dd\vr \,  |\rho_0(\vR; \vr)|^2 
    W_S^{\mathrm{LO}}(m,r) [1-f(r)]
\end{equation}
contains the whole regulator dependence. 
As $[1-f(r)]$ has a very short range, the integrals in $W_0^\text{reg}$ can easily be carried out numerically even for very large Yukawa interaction ranges. 
To tackle $W_0^\infty$, which is for small $m$ (especially for full-square DMEs) much harder to calculate numerically, we split it further according to 
\begin{align}
    W_0^\infty
    &= -\frac{9}{8} \frac{m_\pi^2}{12\pi} \left( \frac{g_{A}}{2 F_\pi} \right)^2 \int \! \dd\vR \sum_{a,b=\mathrm{n,p}} \biggl\{ 
    I_{00} (m,k_a,k_b) \rho_a(\vR) \rho_b(\vR) \nonumber \\
    &\quad \null + I_{02} (m,k_a,k_b) \rho_a(\vR) \left[\frac{1}{4} \vnabla^2 \rho_b(\vR) - \tau_b(\vR) + \frac{3}{5} k_b^2 \rho_b(\vR)\right] \nonumber \\
    &\quad \null + I_{22} (m,k_a,k_b) \left[\frac{1}{4} \vnabla^2 \rho_a(\vR) - \tau_a(\vR) + \frac{3}{5} k_a^2 \rho_a(\vR)\right] \left[\frac{1}{4} \vnabla^2 \rho_b(\vR) - \tau_b(\vR) + \frac{3}{5} k_b^2 \rho_b(\vR)\right]
    \biggr\}
    \,,
\end{align}
where the momentum scales $k_a$, $k_b$ can be set either to the individual momentum scales for neutrons and protons or to the isoscalar momentum scale for both species and the $I_{ij}$ functions depend on the considered DME variant. Zeroth-order DMEs have contributions only from $I_{00}$, truncated-square second-order DMEs have additional contributions from $I_{02}$, and $I_{22}$ contributes only for full-square variants.

The $I_{ij}$ functions are calculated analytically by evaluating the integrals 
\begin{align}
I_{00} (m,k_a,k_b) &\equiv 4\pi \int \! \dd r\, r^2 \Pi_0(k_a r) \Pi_0(k_b r) \frac{\ee^{-m r}}{r} \,, \\
I_{02} (m,k_a,k_b) &\equiv \frac{4\pi}{3} \int \! \dd r\, r^4 \Pi_0(k_a r) \Pi_2(k_b r) \frac{\ee^{-m r}}{r} \,, \\
I_{22} (m,k_a,k_b) &\equiv \frac{4\pi}{36} \int \! \dd r\, r^6 \Pi_2(k_a r) \Pi_2(k_b r) \frac{\ee^{-m r}}{r} \,.
\end{align}
To this end we apply the method outlined in \rcite{Fabr12BesInt} and obtain after some analytical simplifications the following expressions. They are checked against their numerical counterparts for different values of $m$, $k_a$, and $k_b$. 
For the Slater approximation [or any other DME that uses $\Pi_0(x) = 3 j_1(x)/x$] the $I_{00}$ function reads
\begin{align}
    I_{00} (m,k_a,k_b) &= 
    \frac{3 \pi}{4 k_a^3 k_b^3} \biggl\{ 2 k_a k_b \left[ 3\left(k_a^2+k_b^2\right)  -m^2\right] +\left[ -3 \left(k_a^2-k_b^2\right)^2 +6 m^2 \left(k_a^2+k_b^2\right) +m^4\right] \artanh \left(\frac{2 k_a k_b}{k_a^2+k_b^2+m^2}\right) \nonumber \\
    &\quad \null +8 m\left[ \left(k_a^3-k_b^3\right) \arctan
    \left(\frac{k_a-k_b}{m}\right) - \left(k_a^3+k_b^3\right) \arctan \left(\frac{k_a+k_b}{m}\right) \right] \biggr\} \,.
\end{align}
The other $I_{ij}$ functions for NV$^{(2)}$-DME read
\begin{align}
    I_{02} (m,k_a,k_b) &= 
    -\frac{35 \pi}{48 k_a^3 k_b^7} \biggl\{
    4 k_a k_b \left[
        -22 k_a^2 k_b^2 +15 k_a^4 +3 k_b^4
        +6m^2 \left( -13 k_a^2 + k_b^2 \right)
        +3m^4
    \right] 
    \nonumber \\ &\quad \null 
    -6 \left[
        \left(k_a^2 - k_b^2\right)^2 \left(5k_a^2 + k_b^2\right)
        +3m^2 \left(-15k_a^4 + 6k_a^2 k_b^2 + k_b^4\right)
        +3m^4 \left(5k_a^2 + k_b^2\right)
        +m^6
    \right]
    \nonumber \\ &\qquad \null \times
    \artanh \left(\frac{2 k_a k_b}{k_a^2+k_b^2+m^2}\right) 
    \nonumber \\ &\quad \null 
    +48 m k_a^3 \left[3 \left(k_a^2 -k_b^2\right) -5 m^2\right]
    \left[\arctan \left(\frac{k_a-k_b}{m}\right) -\arctan \left(\frac{k_a+k_b}{m}\right)\right] \biggr\} 
\,, 
\end{align}
and
\begin{align}
    I_{22} (m,k_a,k_b) &= 
    \frac{175 \pi}{1536 k_a^7 k_b^7} \biggl( 
    4 k_a k_b \bigl\{
        7 \left(k_a^2+k_b^2\right) \left[-22 k_a^2 k_b^2 +15 \left(k_a^4 + k_b^4\right) \right] 
        -m^2 \left[134 k_a^2 k_b^2 +141 \left(k_a^4 + k_b^4\right)\right] 
        \nonumber \\ &\qquad \null
        -69m^4 \left(k_a^2+k_b^2\right) -15 m^6
    \bigr\} 
    \nonumber \\ &\quad \null
    +6 \Bigl\{
        -7 \left(k_a^2-k_b^2\right)^2 \left[6 k_a^2 k_b^2 + 5 \left(k_a^4+k_b^4\right) \right]
        +28m^2 \left[5 \left(k_a^6+k_b^6\right) +3 \left(k_a^4 k_b^2 + k_a^2 k_b^4\right) \right]
        \nonumber \\ &\qquad \null
        +14m^4 \left[6 k_a^2 k_b^2 + 5 \left(k_a^4+k_b^4\right) \right] 
        +28m^6 \left(k_a^2+k_b^2\right) 
        +5 m^8
    \Bigr\} 
    \artanh \left(\frac{2 k_a k_b}{k_a^2+k_b^2+m^2}\right) 
    \nonumber \\ &\quad \null
    +768 m \left[\left(k_a^7-k_b^7\right) \arctan
        \left(\frac{k_a-k_b}{m}\right)-\left(k_a^7+k_b^7\right) \arctan
        \left(\frac{k_a+k_b}{m}\right)\right] 
    \biggr) \,.
\end{align}

Note that for the case of a single isoscalar momentum scale, i.e., the special case $k_a=k_b$, a Mathematica package to obtain these expressions was published in \rcite{Gebr100BesInt}. For $k_a=k_b$, our equations agree with the ones outputted by that package.

\end{document}